\documentclass{aa}  
\usepackage{graphicx}
\usepackage{natbib}
\usepackage{txfonts}
\usepackage[section]{placeins}
 
\usepackage{hyperref}
\usepackage{caption}

\begin{document}

   \title{Interstellar medium phases and abundances in the central parsec}

   \subtitle{A JWST MIRI/MRS view of the Galactic center}

   \author{P. Vermot \inst{1}, 
          A. Ciurlo \inst{2}, 
          D. Rouan \inst{1}, 
          M.R. Morris \inst{2}, 
          E. Bron \inst{3}, 
          J. Le Bourlot \inst{3}\fnmsep\inst{4}, 
          F. Le Petit \inst{3}, 
          J. Qiu \inst{2},
          A. Togi\inst{4},
          A. Ghez\inst{2}
          T. Do\inst{2}
          \and
          J.R. Lu\inst{5}
          }
   \institute{LIRA, Observatoire de Paris, PSL Research University, CNRS, Sorbonne Université, Université Paris Cité, 5 place Jules Janssen, 92195 Meudon, France
         \and
             Department of Physics and Astronomy, UCLA, Los Angeles, CA 90095-1547,
        \and
         LUX, Observatoire de Paris, PSL Research University, CNRS, Sorbonne Universités, 75014 Paris, France.
         \and
         Université Paris-Cité
         \and
         Department of Physics, 601 University Dr., Texas State University, San Marcos, TX 78666, USA
         \and Department of Astronomy, University of California, Berkeley, USA
             }

   \date{}

    \abstract
   {The Galactic center (GC) is a unique and extreme astrophysical laboratory for studying the interplay between gas, stars, and a supermassive black hole (SMBH). In particular, the circumnuclear disk (CND) and its central cavity (CC) present two contrasting environments in terms of gas content, density, and stellar activity, making them ideal regions in which to study the multiphase structure and chemical composition of the interstellar medium (ISM).}
   {We aim to determine the properties (temperature, density, abundances, and spatial distribution) of the various phases of the ISM in the central parsec of the GC, with particular attention in this paper to the ionized medium.}
   {We used newly obtained observations from the Mid-Infrared Instrument (MIRI) equipped with the Medium Resolution Spectrometer (MRS) aboard the James Webb Space Telescope (JWST) to extract spectra covering the entire spectral range from 5 to 27~$\mu$m in the CND and in the CC. We used the photoionization code CLOUDY to generate synthetic spectra with the same spectral range and resolution, simulating a wide range of gas phases and abundances. We then determined the contribution of each phase to the spectra. Once the abundances and contribution from each phase of the gas were determined, we identified four dominant phases and performed a spatial analysis to determine their contribution to each spaxel of the datacubes.}
  {We find that in both the CND and the CC, the bulk of the emission originates from warm ionized gas with temperatures of between $10^4$ and $10^{4.8}$~K. In the CND, molecular gas contributes significantly to the flux and is spatially structured, while the CC shows minimal molecular gas content, as is expected from these regions. Coronal gas is detected in both regions at the interface between molecular and warm ionized gas. The hottest coronal phase appears faint and patchy in the CC, and has an elongated morphology in the CND. Abundance fitting (in solar-normalized logarithmic units) is primarily constrained by relative abundances: we measure a robust depletion of Fe relative to $\alpha$ elements with $\log(\mathrm{Fe}/\alpha) = -0.78 \pm 0.20$ (CC) and $-0.84 \pm 0.26$ (CND), while CNO is only mildly enhanced relative to $\alpha$, $\log(\mathrm{CNO}/\alpha) = 0.27 \pm 0.20$ (CC) and $0.05 \pm 0.26$ (CND). Absolute abundances are supersolar but more degenerate; the best-fitting models yield $(\log\alpha,\log\mathrm{CNO},\log\mathrm{Fe})=(1.4,1.4,0.4)$ in the CC and $(2.0,1.8,1.2)$ in the CND.}
  {The observed abundance pattern (enhanced CNO and $\alpha$ elements with suppressed Fe) indicates a chemically young environment, recently enriched by core-collapse supernovae and stellar winds, with a limited contribution from older Type Ia supernovae. This favors a scenario of massive, recent star formation rather than cumulative long-term enrichment. Additionally, the projected orientation of the newly identified CND elongated hot coronal feature, perpendicular to the direction toward the SMBH, suggests the action of a large-scale shock possibly resulting from past energetic outflows. }  

\titlerunning{GC-ISM: Ionized gas}
\authorrunning{P. Vermot et al.}
\maketitle
\section{Introduction}

\begin{figure*}
    \centering
    \includegraphics[width=0.95\linewidth]{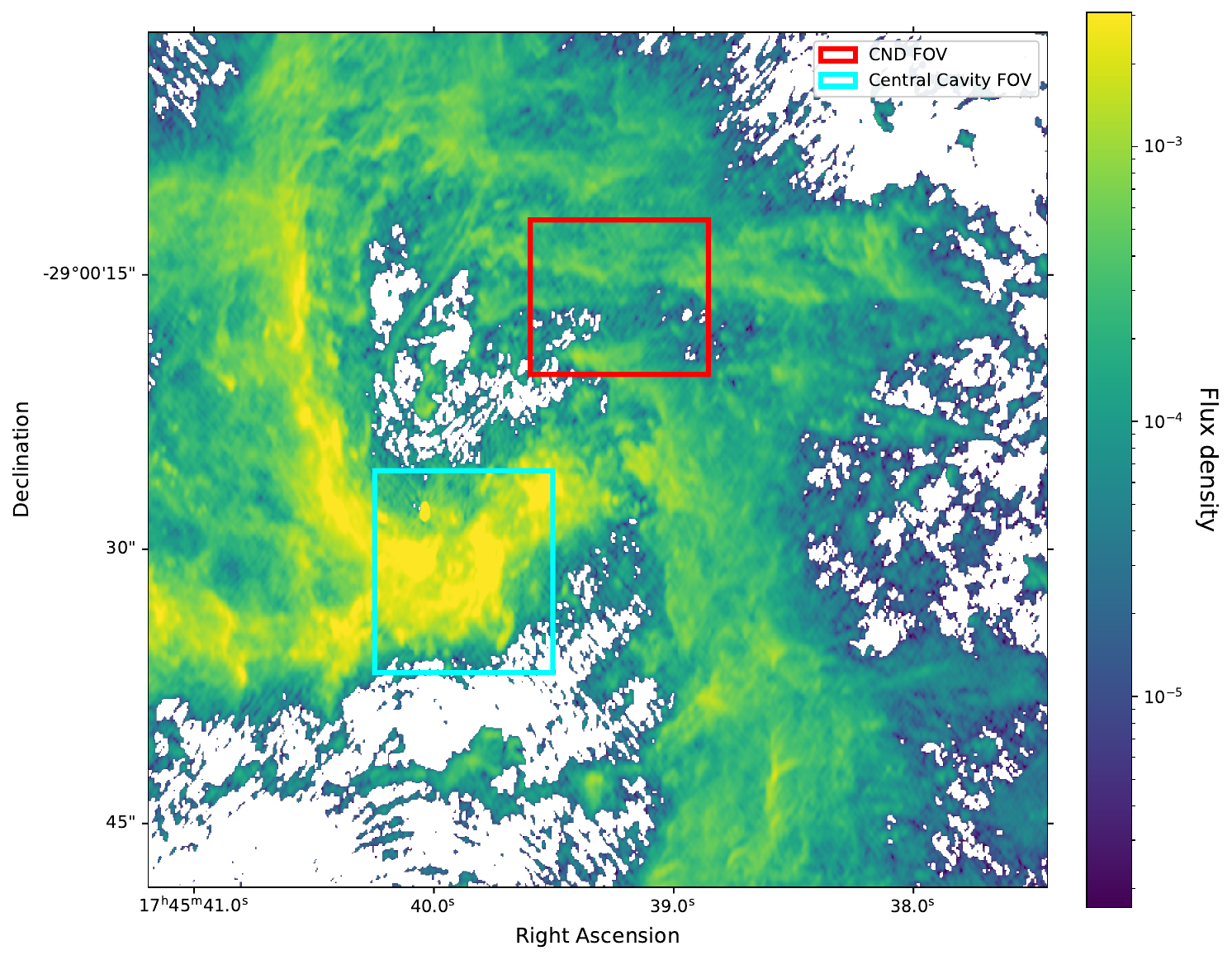}
    \caption{Overview of the GC with the JWST/MIRI-MRS pointings overlaid. The background image is a VLA 2 cm radio map \citep{Morris2017} showing the ionized gas structures in the central few parsecs. The cyan rectangle outlines the CC pointing, and the red rectangle shows the footprint of the CND pointing.}
    \label{fig:galactic_center_overview}
\end{figure*}

\begin{figure*}
    \centering
    \includegraphics[width=\linewidth]{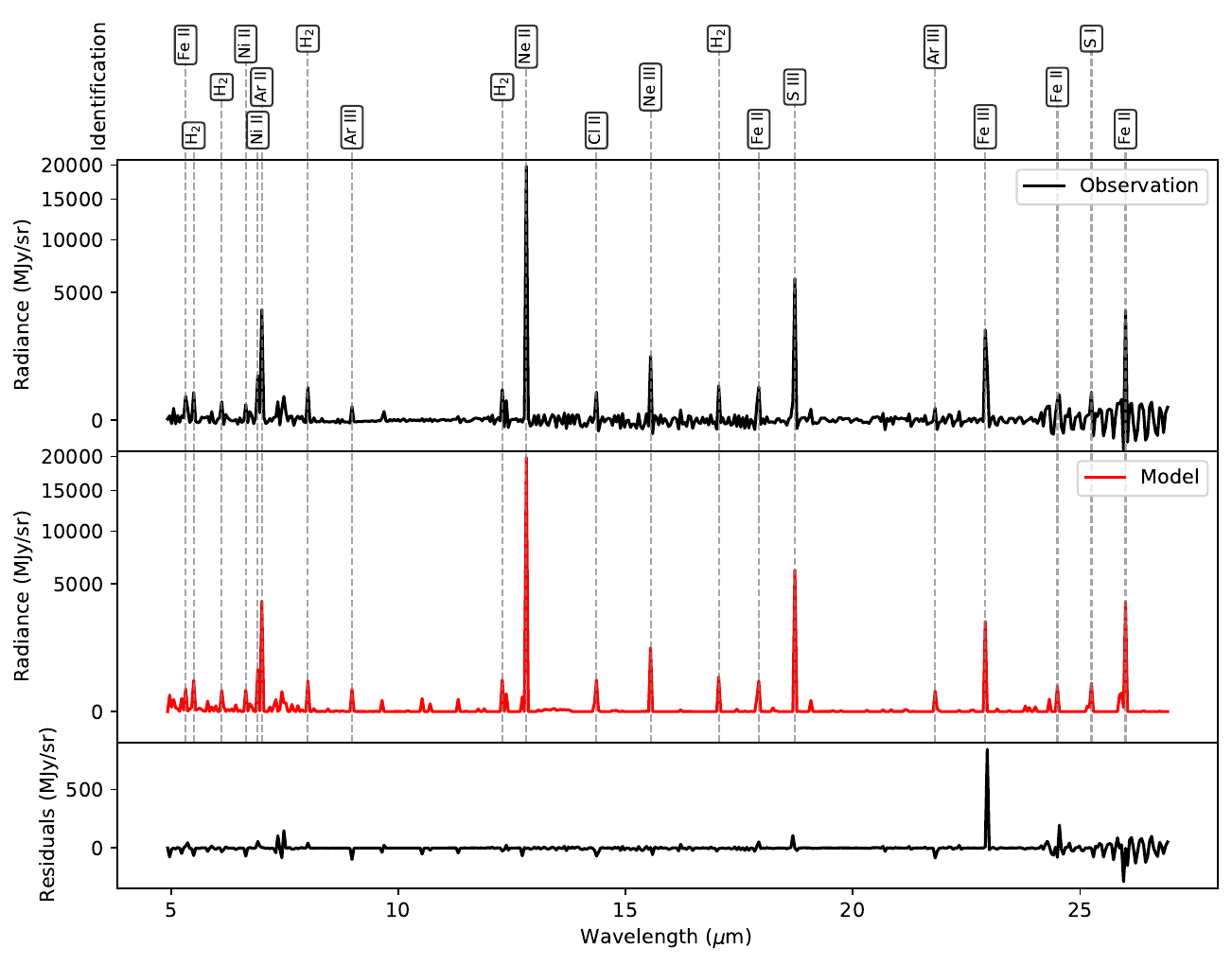}
    \caption{Top: Observed CND spectrum. Middle: Best-fit multiphase model spectrum (red). Bottom: Residuals (observation minus model) as a function of wavelength. The observed spectrum is extracted over the whole CND field of view (extraction and processing in Section~\ref{subsec:data_processing}); the model is the best-fitting simulation from Section~\ref{subsec:composite_and_abundance}, using the abundances of Table~\ref{tab:abundances} and the phase weights of Fig.~\ref{fig:CND_weights}. Line identifications are shown above the model panel: the most significant lines predicted by the model are marked (see Section~\ref{subsec:line_id}); when multiple transitions fall within the same spectral bin, only the brightest is labeled here. The complete identification is provided in Table~\ref{tab:line_identificationcnd}. The y axes of the observation and model spectra are shown on a square-root scale to enhance the visibility of faint features. The two curves are nearly indistinguishable over most of the range; a zoom highlighting fainter features is shown in Fig.~\ref{fig:CND_detailed_comparison_spectra} (Appendix~\ref{app:appendix_multi}).}
    \label{fig:CND_comparison_spectra}
\end{figure*}

\begin{figure*}
    \centering
    \includegraphics[width=\linewidth]{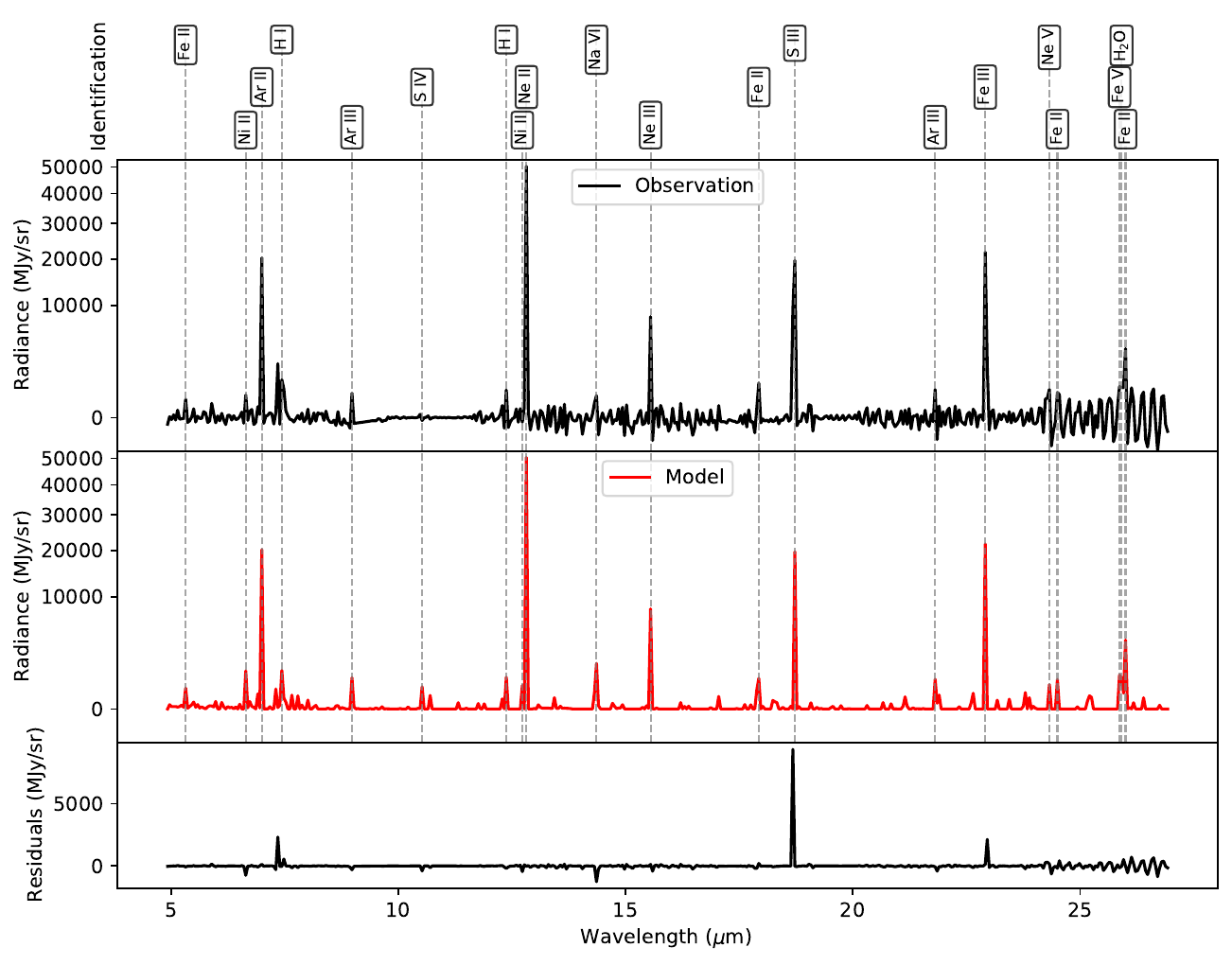}
    \caption{
    Same as Fig.~\ref{fig:CND_comparison_spectra}, but for the CC. The corresponding weights and zoomed-in versions are shown in Figs.~\ref{fig:CC_weights} and~\ref{fig:CC_detailed_comparison_spectra}, and the complete identification in Table~\ref{tab:line_identificationcc}}
    \label{fig:CC_comparison_spectra}
\end{figure*}

The Galactic center (GC) is the closest ($\sim$ 8~kpc) galactic nucleus and, as such, provides a unique laboratory for studying the interplay between stars, gas, magnetic fields, and the central supermassive black hole (SMBH), Sgr~A* \citep{Genzel2003, Ghez2005}. This is known as one of the most extreme regions in the Galaxy, characterized by an exceptionally high stellar density and radiation field, strong turbulence and complex dynamics, and strong magnetic and gravitational fields \citep{Morris1996,Ponti2015}. More broadly, studying the GC improves our understanding of galactic nuclei in general, particularly regarding nuclear star formation, feedback processes, and SMBH fueling.

In this study, we focus on two regions of interest within the GC: the CND and its central cavity (hereafter referred to as CC). The CND is a dense molecular gas ring surrounding the SMBH, with an inner radius of about 1~pc and extending up to an outer radius of about 5~pc. While early virial mass estimates suggested a total mass of several $10^5$~M$\odot$ \citep[e.g.,][]{Christopher2005, MonteroCasta2009}, more recent measurements based on dust continuum emission point to a lower mass of several $10^4$~M$\odot$ \citep{Dinh2021}, suggesting that the clumps are transient and shaped by instabilities rather than self-gravity \citep{Lau2013, Blank2016, Solanki2023}. The region it encircles, the CC, is largely depleted of molecular material \citep[though some remains,][]{Ciurlo2016,Hsieh2019}, and hosts the SMBH Sgr~A* and the compact nuclear star cluster (NSC). The NSC’s winds and intense UV radiation help keep gas scarce \citep{Christopher2005,Genzel2010}, but wind–rim interactions at the CND’s inner edge can also drive instabilities that tend to shrink the cavity \citep{Blank2016}. The CC may further be a downstream result of episodic disturbances \citep[supernovae among the massive YNC stars that stir the CND and promote inward gas migration][]{Dinh2021} and of an earlier phase when the cluster formed and Sgr~A* was more strongly fed, consistent with a limit-cycle picture for GC activity \citep{Dinh2021,Morris1996,Genzel2010}. These two neighboring regions thus present strong contrasts, allowing for a detailed comparative analysis.

In this extreme region of the GC, gas is found in all its forms, from the dense molecular phase to the very hot and diffuse coronal stage, passing through the warm ionized medium \citep{Morris1996, Wang2013}. Studying the temperature and density distribution is crucial for determining the physical state and ionizing mechanisms of the interstellar medium (ISM). Additionally, detailed modeling of the emission lines can help constrain the elemental abundances in the GC, which serve as excellent tracers of the star formation history and enrichment processes \citep{Henry2000, Matteucci2012}.

The James Webb Space Telescope (JWST) Mid-Infrared Instrument (MIRI) provides integral field spectroscopy with its Medium-Resolution Spectrometer (MRS), covering, through its four channels, the entire spectral range from 5~µm to 27~µm with a spectral resolution of R$\geq$1500 and an angular resolution better than 1~$\arcsec$ \citep[the lowest spectral and spatial resolution being at 27$\mu$m][]{Wells2015, Argyriou2023}. This very wide spectral range allows for the observation of numerous emission lines from various elements, probing a broad range of energy levels and densities. Beyond these spectral capabilities, MRS enables the study of spatial variations, making it a powerful tool for investigating the ionized phase of the ISM in this region \citep{AlonsoHerrero2024, CostaSouza2024}.

This work is the first part of a series of papers on the different emitters in the GC. Here, we derive a detailed description of the ISM state in the CND and the center of the CC. To achieve this, we measure the electron density and temperature in the two regions and map the four dominant phases of the ISM: molecular, warm ionized, warm coronal, and hot coronal gas. In Section \ref{sec:obs}, we present the two observations and the post-processing steps performed. In Section \ref{sec:modeling}, we use the photoionization code CLOUDY to quantitatively interpret the physical conditions traced by the emission lines and derive the ISM physical properties, which are presented in Section \ref{sec:results} and discussed in Section \ref{sec:disc}.

\section{Observations, data processing and preliminary analysis}
\label{sec:obs}
\subsection{Observations}

This study is based on two JWST MIRI/MRS observations of the GC, obtained as part of program JWST Cycle 2 GO 3166 (PI: A. Ciurlo). The CND and the CC (see Fig. \ref{fig:galactic_center_overview} were observed on April 18, 2024 and September 11, 2024, respectively, accompanied by two dedicated off-source background exposures (at the same coordinates : RA=17:44:46.647; $\delta$=-28:53:44.41) acquired with identical instrumental setup. Both targets were observed using the full four-channel mosaic mode of the Medium Resolution Spectrometer (MRS), providing full coverage of the $5$–$27\ \mu$m range with spectral resolution $R \sim 3000$–$1500$. Further details on the overall observing strategy for the GO 3166 program, including target selection and calibration rationale, are discussed in Ciurlo et al. (in prep).

\subsection{Data processing}
\label{subsec:data_processing}

Our data processing is detailed in Ciurlo et al. (in prep) and delivers for each field a regularized, continuum-subtracted data cube containing only emission lines, with spatially and spectrally uniform characteristics across the full 5–27~$\mu$m range.
We obtained this final data cube by merging the four channels of each observation \footnote{reduced with Level 3 calibrated products from the Science Calibration Pipeline using pipeline versions 1.13.3 plus mitigation residual detector fringes due to Fabry-Pérot interference \citep{Gasman2023} using \texttt{fit\_residual\_fringes\_1d}} into a single datacube with uniform spatial resolution. 
Each target cube was background-subtracted using the corresponding off-source exposure. The background spectra were extracted by averaging over the background field of view, and they were then subtracted from the target cube.

To isolate the emission line component, we estimated and subtracted the underlying continuum by applying a rolling median filter over 100 spectral bins to each spaxel’s spectrum. This window size was chosen to be significantly broader than the width of typical emission lines (which span only a few bins), while remaining narrower than the large-scale continuum variations. This ensures that the median primarily traces the continuum rather than the lines. Tests on selected spaxels confirmed that any residual bias remains negligible compared to the line fluxes. The procedure effectively preserves line shapes and intensities while suppressing low-frequency continuum features.

The background-subtracted and continuum-corrected datacubes were then rebinned to a lower spectral resolution of $\Delta \lambda = 0.04~\mu\mathrm{m}$ to improve signal-to-noise and reduce computational load. This was performed using fractional rebinning, i.e., an exact averaging method that preserves integrals even for non-integer rebinning factors, and therefore free of interpolation artifacts. In this approach, each output bin receives a weighted average of all input pixels that overlap it, with weights given by the fractional overlap. In practice, the original spectrum length of $11637$ points was reduced to $500$, corresponding to a rebinning factor of $\sim 23$. Integrated spectra over the Channel~1 field of view were extracted and are shown in Figs.~\ref{fig:CND_comparison_spectra} and \ref{fig:CC_comparison_spectra}, forming the basis of the analysis presented here. Spatial variations within the cubes are explored in Section~\ref{subsec:spatial}. More comprehensive documentation of the common processing strategy for all GO 3166 observations will be provided in Ciurlo et al. (in prep).

\section{Modeling}
\label{sec:modeling}
To interpret the physical conditions of the ionized gas traced by the emission lines, we used the photoionization code CLOUDY \citep{Ferland2017}. CLOUDY self-consistently predicts the intensities of a wide range of emission lines based on the gas density, temperature, and elemental abundances. By comparing these to our observations, we can constrain properties such as electron temperature, density, and abundances. This approach enables a quantitative analysis of the ionized ISM in both the CND and CC.

We initially focus on the analysis of spatially integrated spectra extracted from the two observed fields (CND and CC). To determine the physical conditions in these two regions of interest, we employed a multistep modeling approach. We first generated a large grid of CLOUDY simulations spanning a wide range of temperatures, densities, and elemental abundances \ref{tab:cloudy_params}. Each simulation was used to produce a synthetic mid-infrared spectrum matching our observations in terms of spectral range and spectral resolution. We then compared these models to the observed integrated spectra, initially through one-to-one matching, and later by determining for each abundance bin an optimal multiphase combinations of models. Finally, we used the best multiphase model to generate four spectral templates corresponding to the main ISM phases, and applied them to extract spatially resolved information across the fields.

\subsection{Simulation setup}

For each pair of temperature and density (T,n) within the ranges listed in Table~\ref{tab:cloudy_params}, we computed a single-zone CLOUDY simulation of a gas cloud in thermal equilibrium. These simulations include background heating from the cosmic microwave background and galactic cosmic rays, but do not incorporate any explicit external radiation field (e.g., UV illumination). This choice does not imply the absence of radiation in the GC--which is known to host strong radiation fields--but rather reflects a modeling strategy in which temperature is treated as a controlled input parameter, allowing us to explore its effect on the emitted spectrum without modeling the exact heating mechanisms. The simulations include ionized, atomic, and molecular processes, though molecular species only contribute significantly at the lowest temperatures and highest densities. Each simulation uses a static, uniform gas layer, providing a local emissivity spectrum for a given set of physical conditions. While this simplification neglects gradients and non-equilibrium effects, it enables a robust and flexible fit to the observed emission lines, and yields meaningful constraints on the thermal and chemical structure of the emitting gas.

For each cloud, the output is the intensity of a large number of emission lines, which we converted into a spectrum with the same wavelength range and spectral binning as our datasets. Each set of simulations thus produces 793 spectra (61 temperature bins $\times$ 13 density bins).

We then computed such simulation grids for a wide range of abundances. We split the elements into three groups corresponding to distinct astrophysical processes \citep{Matteucci1986, Woosley1995, Nomoto2013, Karakas2014}: 
\begin{itemize} 
\item[\textbullet] CNO elements: Elements involved in the CNO cycle of massive stars, which are released continuously over time from AGB and Wolf-Rayet stars, and thus trace young and intermediate-age stellar populations and stellar winds. This group includes: C, N, O, Li, Be, B, F, Na, Al, P, Cl, Ar, and K. 
\item[\textbullet] $\alpha$ elements: Synthesized in Type~II (core-collapse) supernovae, these elements trace very massive stars, and therefore very recent star formation. The elements considered in this group are: Ne, Mg, Si, S, Ca, and Ti.
\item[\textbullet] Fe-peak elements: Produced mainly in Type~Ia supernovae originating from older, low-mass binary systems, and thus trace older integrated star formation histories. We include in this group: Fe, Mn, Co, Ni, Cu, Zn, Cr, V, and Sc. \end{itemize}

For each group, we explored a range of abundances relative to solar values, from -2~dex to +2~dex in steps of 0.2~dex. The abundances of the primordial elements H and He were kept fixed at their solar values. In total, this results in approximately 7 million simulations available for comparison with the observations: 9261 abundance triplets, each associated with 793 pairs of (T,n).

For readability, we write $\log X$ and $\log(X/Y)$ in compact form. Unless stated otherwise, $\log$ denotes base 10 and
\[
\log X \equiv \log_{10}\!\left(\frac{X}{X_\odot}\right), \qquad 
\log(X/Y) \equiv \log_{10}\!\left(\frac{X/X_\odot}{Y/Y_\odot}\right),
\]
i.e., abundances are always solar-normalized before taking the logarithm.

\begin{table}
    \caption{Parameters range used for our CLOUDY simulations}
    \centering
    \begin{tabular}{ccccc}
        \hline
        \hline
        Parameter & Unit & Min & Max & Step \\
        \hline
        Temperature & log(K) & 1 & 7 & 0.1\\
        Density & log($cm^{-3}$) & -4 & 8 & 1 \\
        $\alpha$ abundance & log($\frac{\alpha}{\alpha_\odot}$) & -2 & 2 & 0.2 \\
        CNO abundance & log($\frac{\mathrm{CNO}}{\mathrm{CNO}_\odot}$) & -2 & 2 & 0.2 \\
        Fe-peaked abundance & log($\frac{\mathrm{Fe}}{\mathrm{Fe}_\odot}$) & -2 & 2 & 0.2 \\
        \hline
    \end{tabular}
    \label{tab:cloudy_params}
\end{table}

\subsection{Extinction correction}

Throughout the fitting procedure, we applied interstellar extinction to all our model spectra using the extinction curve from \citet{Chiar2006}, adopting a fiducial value of $A_V = 30$. We tested several alternative extinction prescriptions -- including those of \citet{Fritz2011} and \citet{Sanders2022} -- but found that the choice of extinction curve has a limited impact on our analysis of the ionized gas spectrum. However, the preliminary results of a companion study focused on the modeling of molecular hydrogen emission lines (Qiu et al., in prep.) show that the extinction curve from \citet{Chiar2006} provides the best fit among these extinction prescriptions to H$_2$ excitation diagrams, particularly in reproducing the flux ratios of the S(1) and S(2) lines. This improvement stems from the stronger silicate absorption feature around 18~$\mu$m in the Chiar et al. curve, which is nearly twice as deep as in other models, and effectively brings the S(1) line into physically consistent alignment with the rest of the excitation diagram.

While the dense sampling of emission lines across our spectral range, in principle, allows for a fit of the extinction parameter \( A_V \) directly from the data, we found that it is strongly degenerate with gas temperature, density, and abundances. Our attempts to fit for extinction yielded large uncertainties and did not improve the fit compared to fixed fiducial values. Given the extensive literature on the GC and the well-established extinction values from dedicated observations \citep[e.g.,][]{Cotera2000, Scoville2003, Schodel2010, Fritz2011, Nogueras2019}, we adopted a fixed value of \( A_V = 30 \), which is commonly used in prior studies of the CND and CC.

In the wavelength range of MIRI/MRS, the dominant effect of extinction is due to the strong silicate absorption bands centered at 10~$\mu$m and 18~$\mu$m. These features affect key diagnostic lines such as [S~III]~$18.7~\mu$m, and their modeling is therefore essential for reliable physical interpretation.

\subsection{Modeling of the full integrated spectra}
\label{subsec:multiphas}
Instead of relying on specific emission line ratios to determine the physical conditions of the emitting medium, we chose to model the entire observed emission line spectrum. This method presents several key advantages compared to standard emission line ratio fitting: it takes into account the flux from all detected emission lines, it incorporates information from non-detections, and it makes no a priori assumptions about line identification. The main drawback is the computational cost, as this approach requires generating a complete spectrum for each simulation.

To mitigate this, we created binned integrated spectra for the CND and CC regions with a much lower spectral resolution ($\Delta \lambda = 0.04~\mu\mathrm{m}$, see \ref{subsec:data_processing}), while still resolving all the main emission lines. Testing our method on a subset of simulations confirmed that this binning did not affect the results. Moreover, this approach offers the added advantage of eliminating the effects of Doppler shifts (up to $\sim 400\ \mathrm{km.s}^{-1}$), which simplifies the comparison between model and observed spectra. However, this comes at the cost of losing information, since potential multi-component emission features at different velocities are blended. We note that this blending is not perfect, and the high velocity contributions contribute significantly to the residuals visible in Figs. \ref{fig:CND_comparison_spectra} and \ref{fig:CC_comparison_spectra}.

\subsubsection{One-to-one comparison}
\label{subsubsec:monophas}
As a first step in fitting the observed spectra, we performed a one-to-one comparison between each synthetic spectrum in our simulation grid and the observed spectrum. For each simulation, we computed a normalized residual defined as

\begin{equation}
R = \sum_i \left( \frac{S^{\mathrm{obs}}_i}{\sum_j S^{\mathrm{obs}}_j} - \frac{S^{\mathrm{mod}}_i}{\sum_j S^{\mathrm{mod}}_j} \right)^2
\end{equation}

where \( S^{\mathrm{obs}}_i \) and \( S^{\mathrm{mod}}_i \) represent the observed and model spectra at wavelength bin \( i \). This metric compares the shapes of the normalized spectra, minimizing sensitivity to overall flux scaling and emphasizing line ratios and spectral features. While this single-phase fitting approach is not expected to capture the full complexity of the emitting medium, it offers a fast and systematic method to explore parameter space and gain preliminary constraints on physical conditions.

\subsubsection{Multiphase fitting}
\label{subsubsec:multiphas}
The second step of our analysis was to fit a linear combination of synthetic spectra to the observed spectrum in order to account for the likely multiphase nature of the ionized gas. For a given abundance combination, we precomputed all 793 spectra $S_k$ (with flux normalized to unity) corresponding to the full $(T, n)$ grid, and searched for the optimal set of non-negative weights, \( w_k \), that the combined model spectrum \( S^{\mathrm{mod}} = \sum_k w_k S_k \) best matches the observed spectrum.

This optimization was carried out using the L-BFGS-B algorithm implemented in the \texttt{minimize} function from \texttt{scipy.optimize} \citep{Zhu1997}. The cost function minimized was the squared residuals between the model and observed normalized spectra:

\begin{equation}
\mathcal{L}(\mathbf{w}) = \sum_i \left( \left( \sum_k w_k S_{k,i} \right) - \frac{S^{\mathrm{obs}}_i}{\sum_j S^{\mathrm{obs}}_j} \right)^2 \mathrm{.}
\end{equation}
Although the number of free parameters (793 weights) exceeds the number of independent spectral bins (500), the effective number of degrees of freedom is considerably lower in practice, because a large fraction of weights are effectively constrained to zero (see Figs.~\ref{fig:CND_weights} and \ref{fig:CC_weights}). This is consistent with the physical expectations that only a limited subset of astrophysical gas phases can contribute significantly to the observed emission line spectrum in this wavelength range.

To ensure that the solution was not a result of local minima, we repeated the minimization five times with different random initializations and retained the solution with the lowest cost function value. Across multiple test cases, we observed that the resulting weight distributions consistently converged to nearly identical solutions. This consistency provides good confidence in the robustness and stability of the fit, despite the apparent high dimensionality of the parameter space. For each abundance triplet, this procedure yields an optimized set of weights characterizing the relative contributions of distinct ISM phases to the observed spectrum.

This fitting procedure was then applied across all abundance bins for the $\alpha$, CNO, and Fe element groups, allowing us to compare the squared residuals obtained for each case. The final product of this process is a 3D cube of squared residuals as a function of $\alpha$, CNO, and Fe abundances, which is presented in Figs.~\ref{fig:CND_abundances} and \ref{fig:CC_abundances} as 2D projections (where each pixel corresponds to the minimum value of the squared residuals found along the third dimension). The results are summarized in Table~\ref{tab:abundances}.

\subsection{Spatial decomposition of the ISM phases}
\label{subsec:spatial_distri}
While our full multiphase spectral fitting approach provides detailed insight into the physical conditions of the ionized gas, it is computationally prohibitive to apply this method to each spaxel of the datacubes. Fitting a single spectrum requires computing the optimal weights for all 793 components in the $(T, n)$ grid, repeated five times per abundance bin to ensure convergence. For all 9261 abundance triplets, this results in several thousands core-hours per integrated spectrum.

To extend our analysis to spatially resolved information, we leveraged the results of the multiphase fit described above by constructing composite spectral templates for different ISM temperature regimes. Specifically, we defined four temperature intervals: 
\begin{itemize}
    \item[\textbullet] molecular gas ($T \leq 10^4\ \mathrm{K}$)
    \item[\textbullet] warm ionized gas ($10^4\ \mathrm{K} \leq T < 10^{4.8}\ \mathrm{K}$)
    \item[\textbullet] warm coronal gas ($10^{4.8}\ \mathrm{K} \leq T < 10^5\ \mathrm{K}$)
    \item[\textbullet] hot coronal gas ($T \geq 10^5\ \mathrm{K}$). 
\end{itemize}
Each phase template was constructed by summing the optimally weighted model spectra corresponding to that temperature range, using the weights derived from the integrated spectrum fit.

For each spaxel in the datacube, we then performed a least-squares fit to determine the optimal linear combination of these four phase templates that best matches the observed spectrum. This approach enables us to construct spatially resolved maps of the relative contribution of each phase to the total line emission, while avoiding the full parameter search in each spaxel.

\section{Results}
\label{sec:results}

\begin{figure*}
    \centering
    \begin{tabular}{ccc}
        \includegraphics[width=0.33\linewidth]{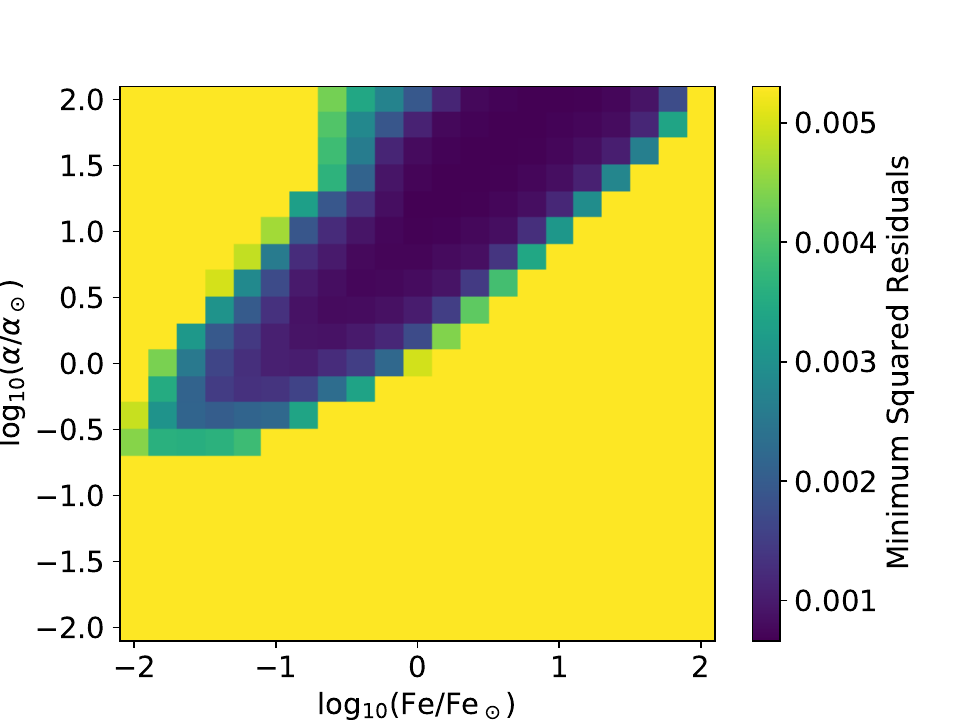} &
        \includegraphics[width=0.33\linewidth]{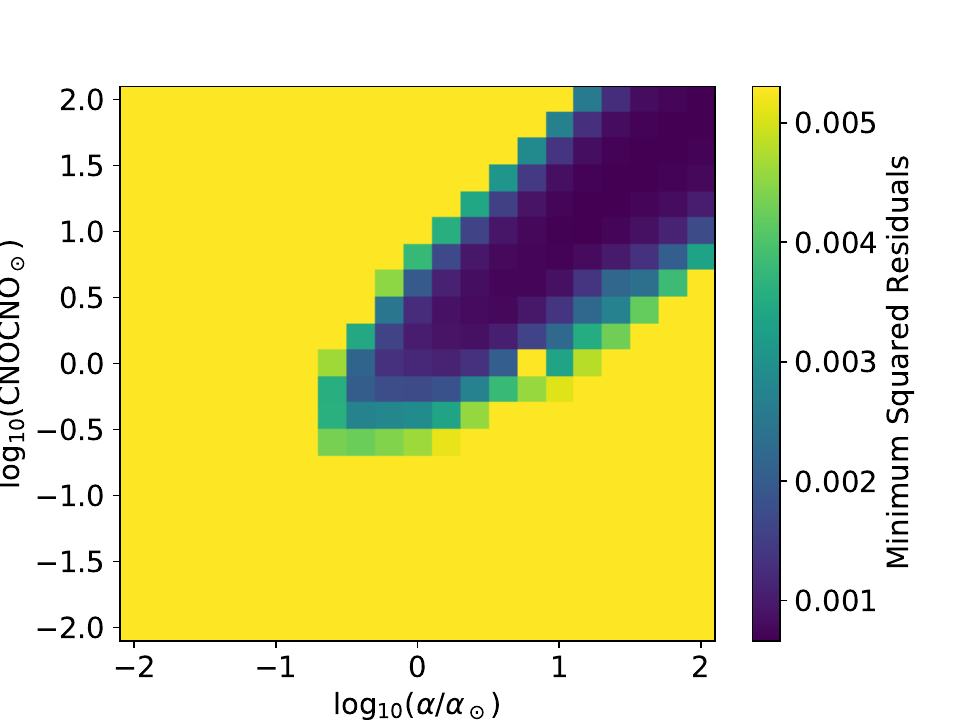} &
        \includegraphics[width=0.33\linewidth]{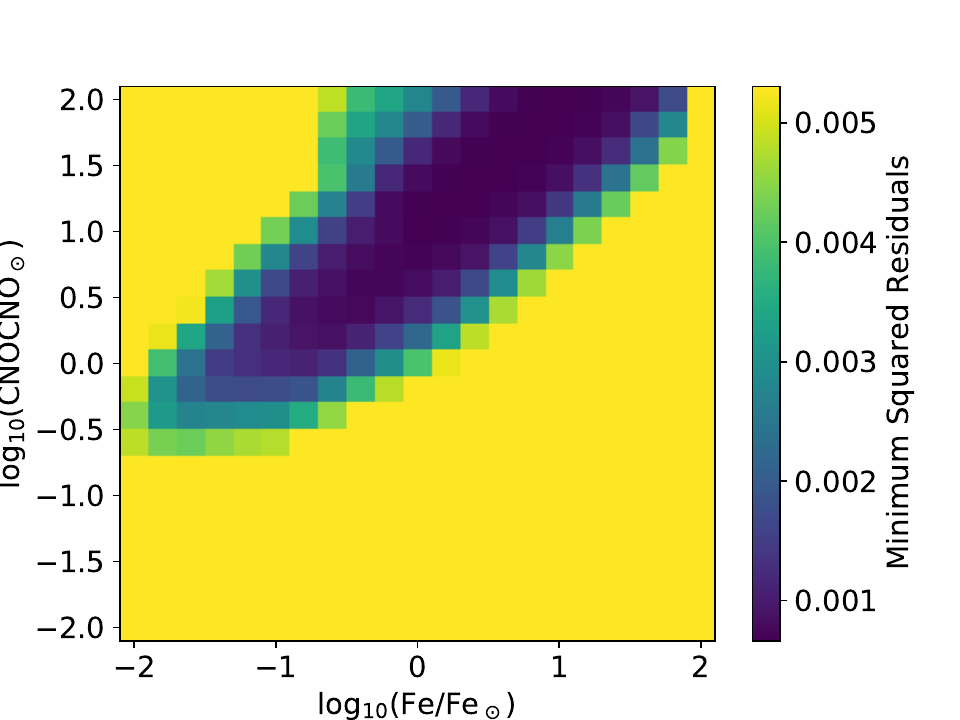} \\
        (a) & (b) & (c) \\
    \end{tabular}
    \caption{
    Residuals as a function of relative abundances in the CND. Panels show 2D projections for (a) $\alpha$/Fe, (b) CNO/$\alpha$, and (c) CNO/Fe. Darker shades indicate better fits. Diagonal valleys suggest overall enrichment and a relative enhancement of CNO and $\alpha$ elements compared to Fe by $\sim$0.5 dex.
    }
    \label{fig:CND_abundances}
\end{figure*}

\begin{figure*}
    \begin{tabular}{ccc}
        \includegraphics[width=0.33\linewidth]{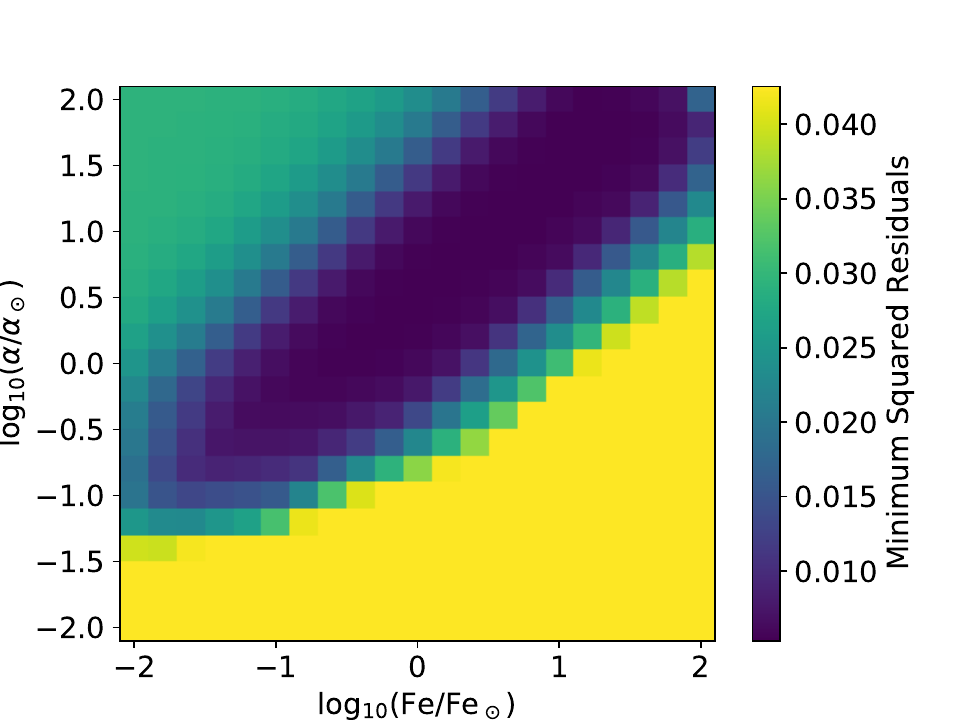} &
        \includegraphics[width=0.33\linewidth]{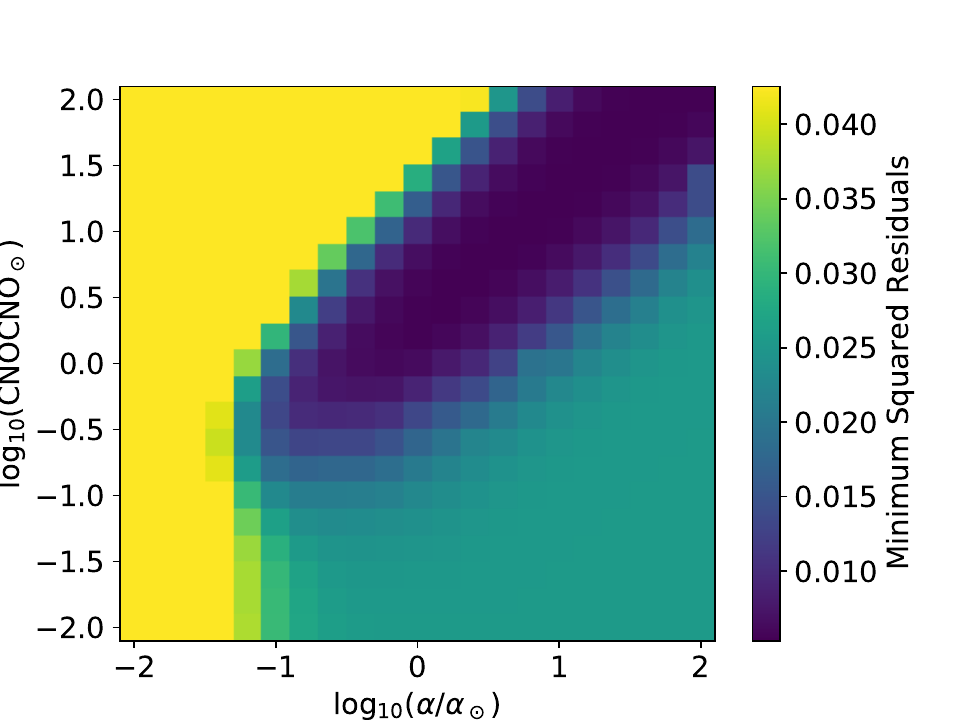} &
        \includegraphics[width=0.33\linewidth]{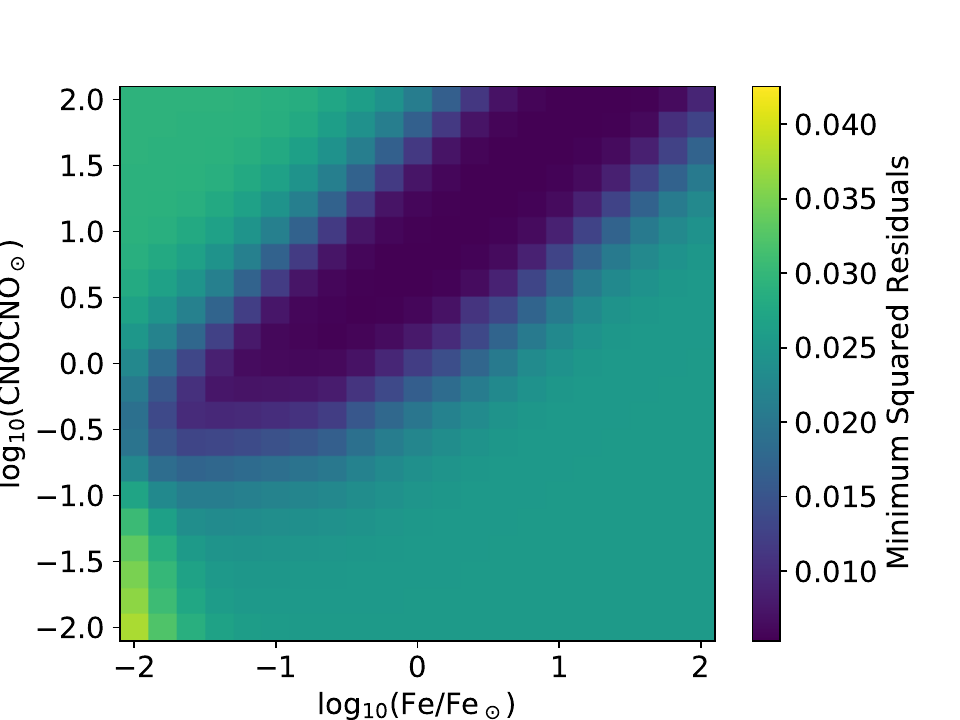} \\
        (a) & (b) & (c) \\
    \end{tabular}
    \caption{Same as Fig.~\ref{fig:CND_abundances}, but for the CC}
    \label{fig:CC_abundances}
\end{figure*}

\begin{figure}
    \centering
    \includegraphics[width=0.99\linewidth]{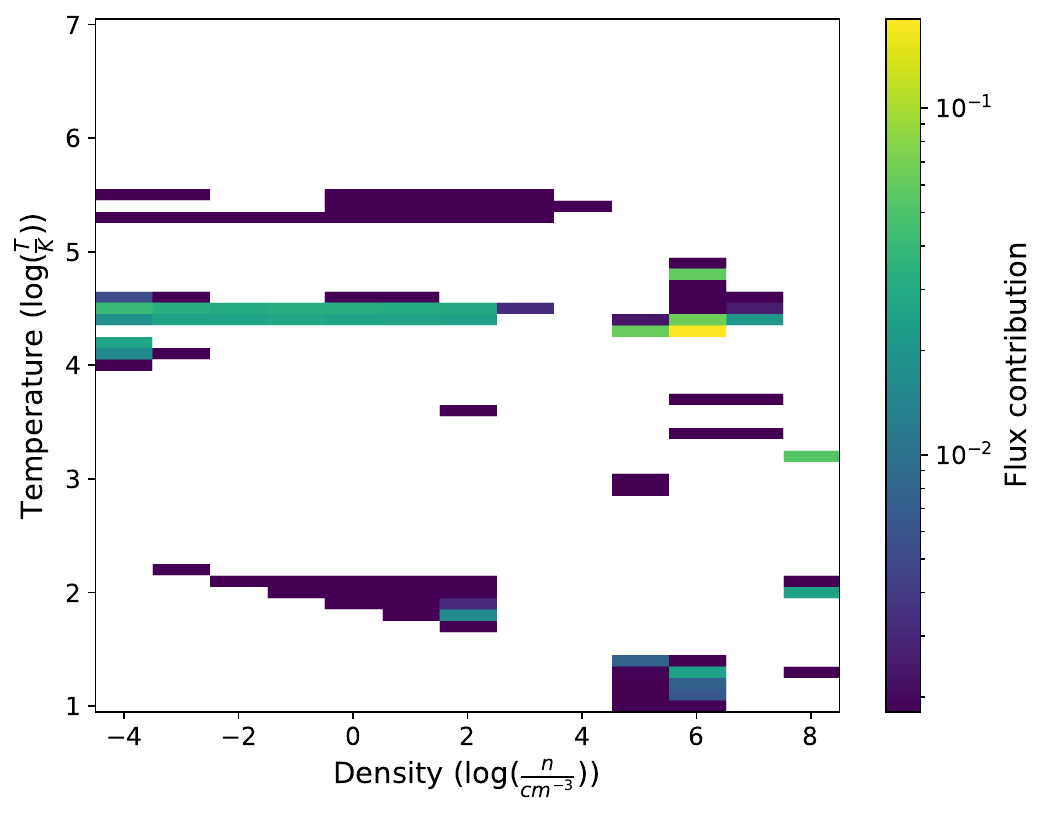}
    \caption{Contribution of several (T,n) gas phases to the CND spectrum in the best-fitting multiphase model (see Section~\ref{subsubsec:multiphas}). The map shows relative flux weights assigned to each phase, with yellow indicating strong contribution, dark blue negligible, and white null. A logarithmic color scale highlights faint components. While warm ionized gas near T $\sim 10^{4.5}$~K dominates, significant cold (molecular) and hot (coronal) components confirm the multiphase nature of the emission.}

    \label{fig:CND_weights}
\end{figure}

\begin{figure}
    \centering
    \includegraphics[width=0.99\linewidth]{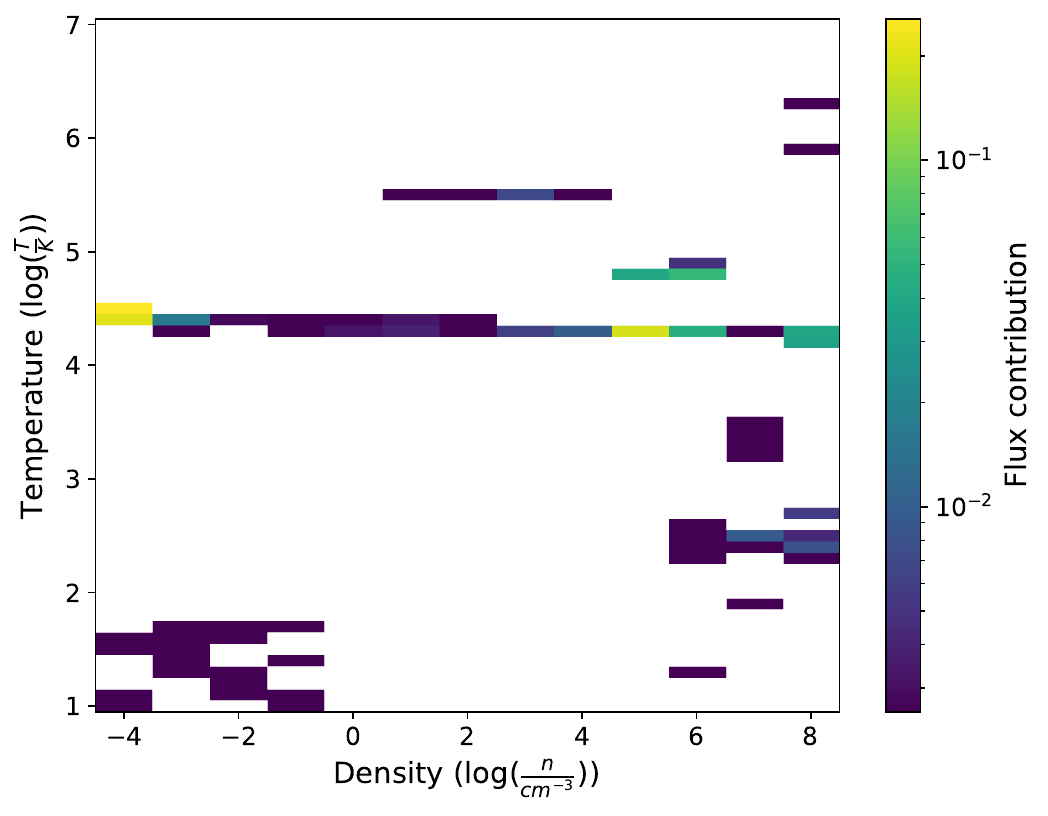}
    \caption{Same as Fig.~\ref{fig:CND_weights}, but for the CC spectrum}
    \label{fig:CC_weights}
\end{figure}

\begin{figure*}
    \centering
    \begin{tabular}{cccc}
        \includegraphics[width=0.24\linewidth]{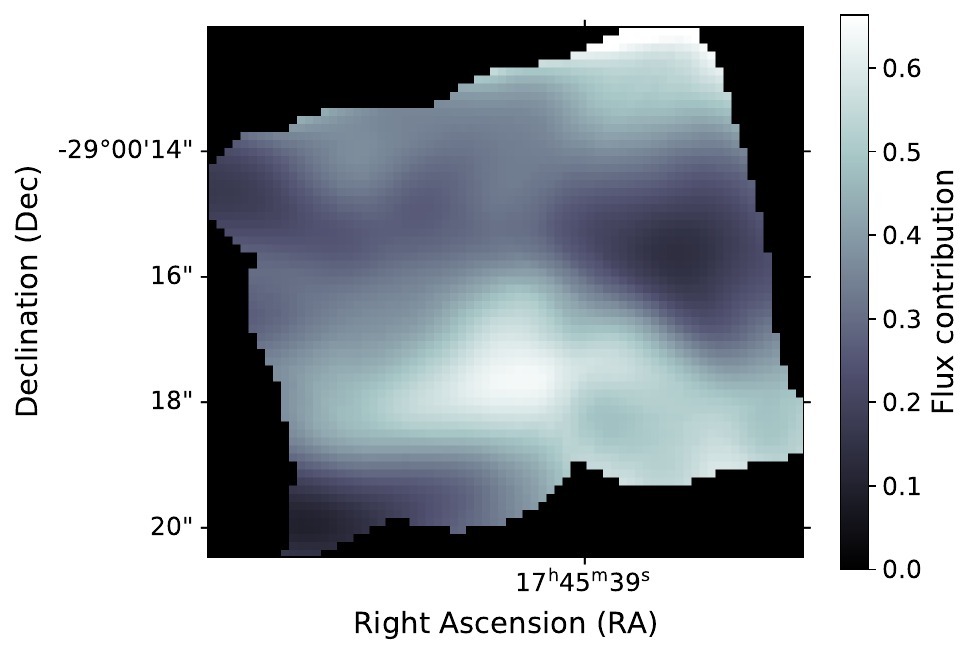} &
        \includegraphics[width=0.24\linewidth]{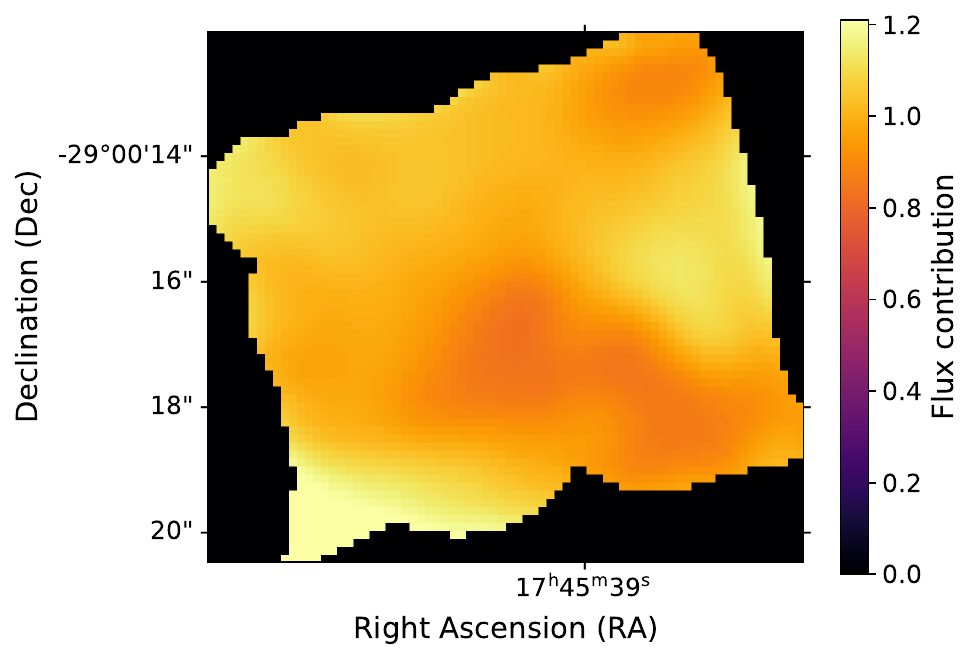} &
        \includegraphics[width=0.24\linewidth]{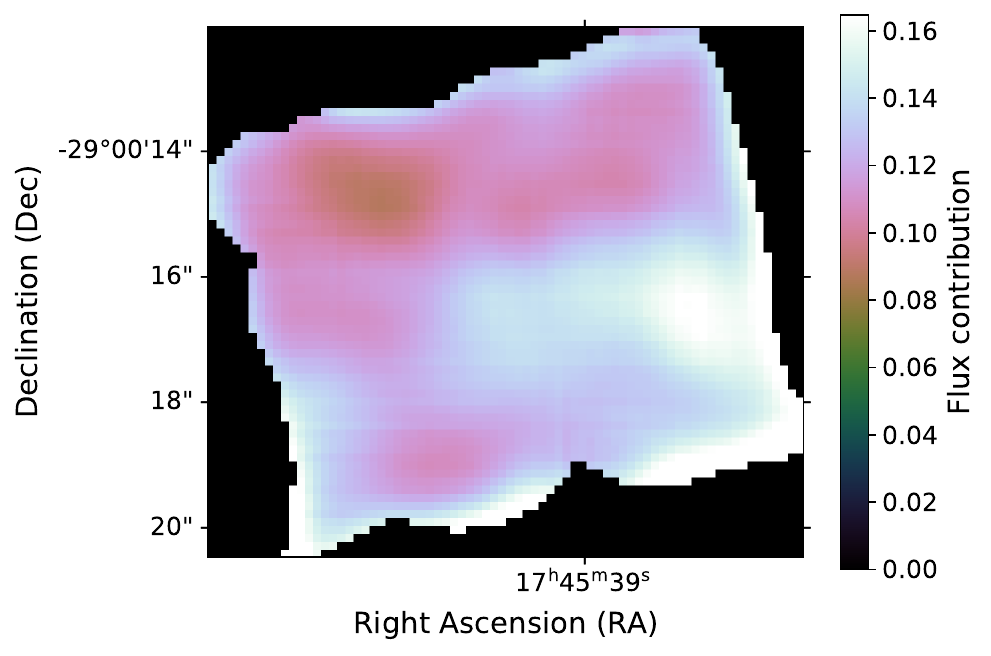} &
        \includegraphics[width=0.24\linewidth]{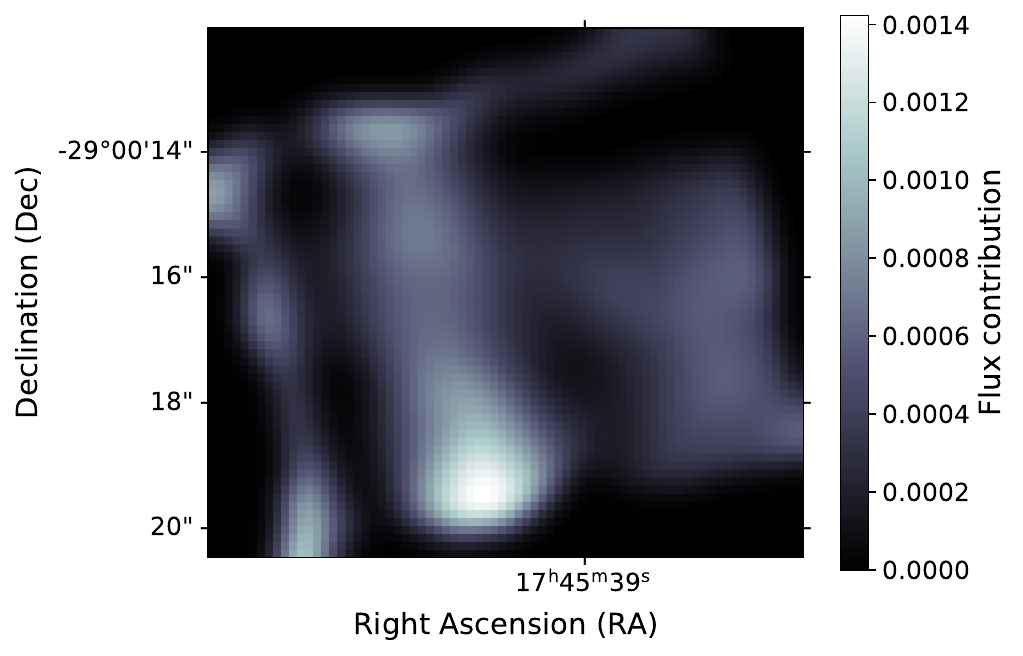} \\
        (a) Cold molecular & (b) Warm ionized & (c) Warm coronal & (d) Hot coronal \\
    \end{tabular}
    \caption{
    Spatial distribution of the four gas phases in the CND. Each panel shows the surface brightness contribution of a distinct phase: (a) cold molecular ($T \leq 10^4$~K), (b) warm ionized ($10^4$–$10^{4.8}$~K), (c) warm coronal ($10^{4.8}$–$10^5$~K), and (d) hot coronal gas ($T \geq 10^5$~K). The maps are derived by fitting the composite spectra described in Section~\ref{subsec:spatial_distri}. See Sect.~\ref{subsec:spatial_distri} for details.
    }
    \label{fig:CND_spatial}
\end{figure*}

\begin{figure*}
    \centering
    \begin{tabular}{cccc}
        \includegraphics[width=0.24\linewidth]{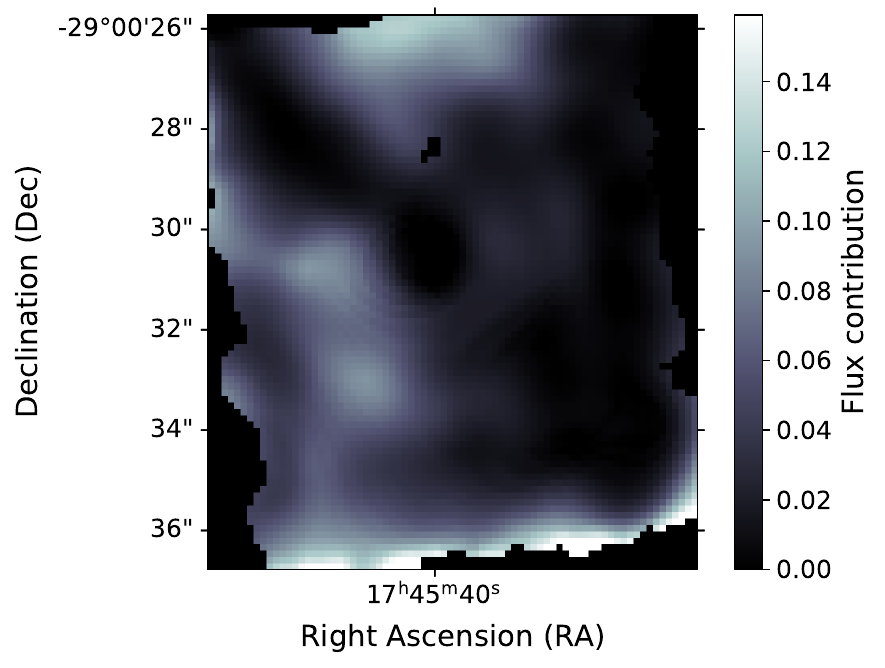} &
        \includegraphics[width=0.24\linewidth]{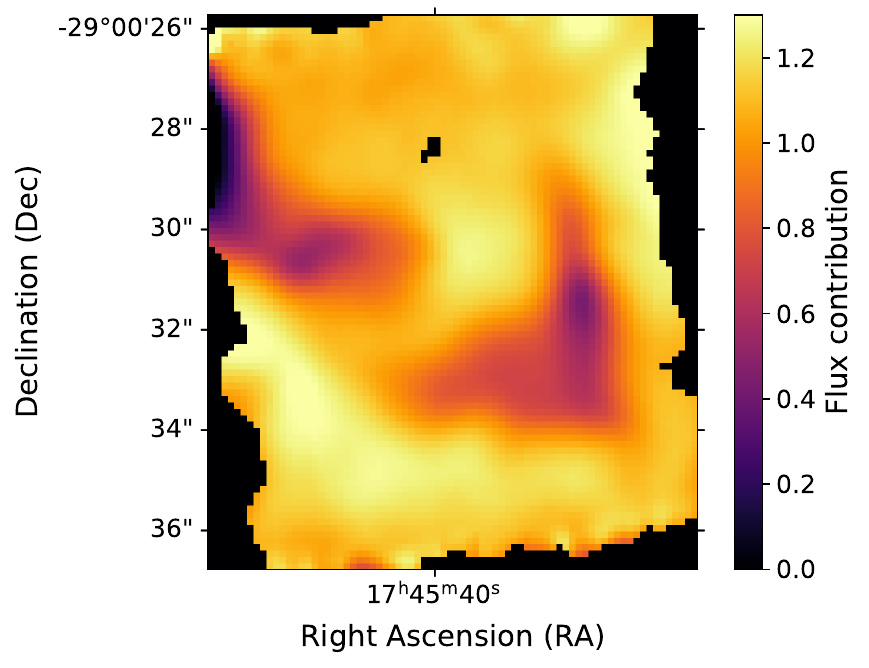} &
        \includegraphics[width=0.24\linewidth]{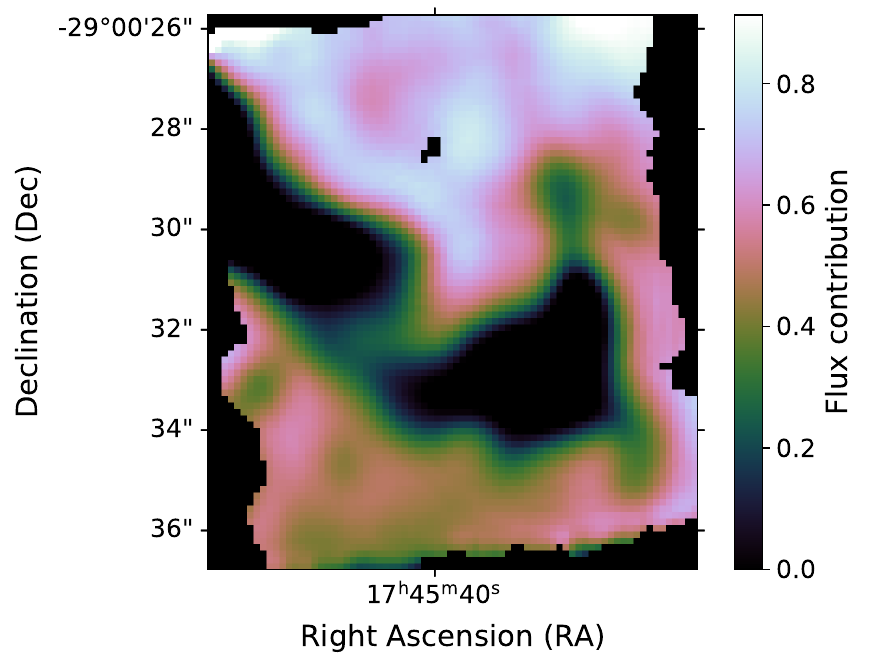} &
        \includegraphics[width=0.24\linewidth]{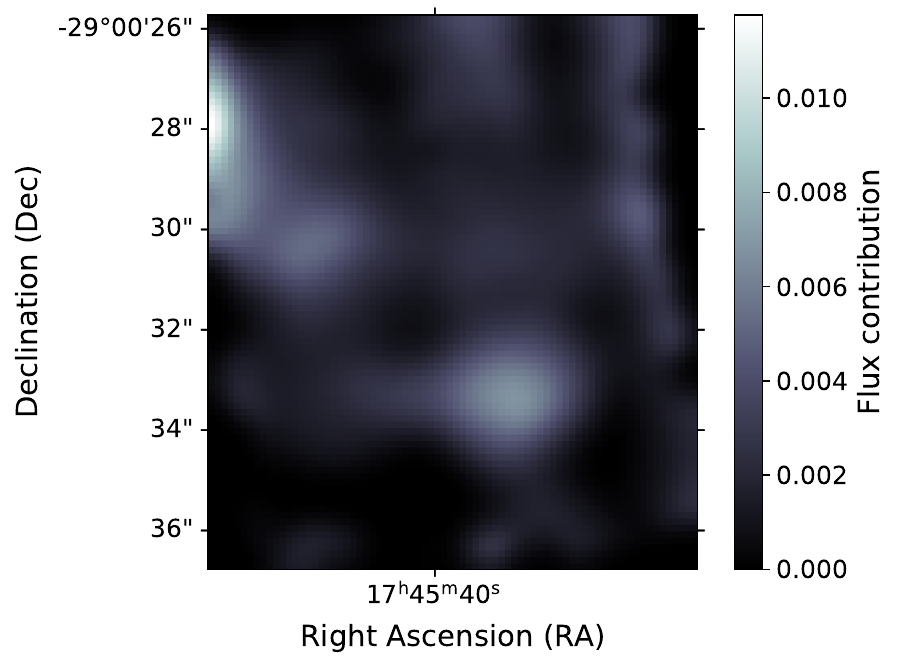} \\
        (a) Cold molecular & (b) Warm ionized & (c) Warm coronal & (d) Hot coronal \\
    \end{tabular}
    \caption{
    Same as Fig.~\ref{fig:CND_spatial}, but for the CC.
    }
    \label{fig:CC_spatial}
\end{figure*}
\subsection{Integrated spectra: a single temperature and density medium model}

As a preliminary step, we tested how well a single-phase gas model could reproduce the observed spectra. Although the fits are significantly worse than with our multiphase model, this analysis brings interesting results, in particular that most of the line emission in both regions originates from warm ionized gas around $10^{4.3}$~K. The full results of this comparison are discussed in Appendix~\ref{app:appendix_residuals}.

\subsection{Integrated spectra: multiphase medium}
\label{subsec:composite_and_abundance}

We then applied the multiphase fitting method described in Section~\ref{subsec:multiphas}, computing the optimal linear combination of synthetic spectra for each abundance triplet. The resulting 3D cube of squared residuals, covering the full grid of $\alpha$, CNO, and Fe abundances, is summarized in Figs.~\ref{fig:CND_abundances} and \ref{fig:CC_abundances} as 2D projections, where each pixel shows the minimum residual found along the third dimension. 

To determine the best fit and uncertainties, we define the optimal solution as the model that yields the global minimum residual (best fit). To quantify degeneracies and estimate uncertainties on relative abundances, we selected all models with squared residuals less than three times this minimum (the “valley” of acceptable solutions) and performed a principal component analysis (PCA) on their abundance vectors $(\log\alpha,\log\mathrm{CNO},\log\mathrm{Fe})$. The first principal component gives the valley’s direction; for interpretability, we normalize this direction by its $\alpha$ component and represent the valley locally as
\[
\vec{P}(s)=\vec{x}_0+s\,\vec{v},
\]
where $\vec{x}_0$ denotes the center of the solution valley (used to determine the abundance ratios) and $\vec{v}$ the normalized direction. The orthogonal dispersion of solutions around this line, measured at $\vec{x}_0$, provides uncertainties on $\log(\mathrm{CNO}/\alpha)$ and $\log(\mathrm{Fe}/\alpha)$. Region-specific best-fit values, ratios, and valley parametrizations are given below and summarized in Table~\ref{tab:abundances}, and illustrated in Figs. \ref{fig:CND_PCA} and \ref{fig:CC_PCA}.

\subsubsection{CND}

For the CND, we find a CNO and $\alpha$ enrichment combined with a relative Fe depletion. The best-constrained quantities are the relative abundances:
\[
\log\!\left(\frac{\mathrm{CNO}}{\alpha}\right) = 0.05 \pm 0.26, \qquad
\log\!\left(\frac{\mathrm{Fe}}{\alpha}\right) = -0.84 \pm 0.26.
\]
The best-fitting model has absolute abundances (dex relative to solar):
\[
\log\alpha = 2.0,\quad \log\mathrm{CNO} = 1.8,\quad \log\mathrm{Fe} = 1.2.
\]
The PCA-derived valley, using a local center solely for parametrization, is
\[
\vec{P}_{\mathrm{CND}}(s) = 
\begin{bmatrix} 1.09 \\ 1.14 \\ 0.25 \end{bmatrix}
+ s \cdot 
\begin{bmatrix} 1.03 \\ 1.00 \\ 1.13 \end{bmatrix},
\]
i.e., for every unit change in $\log(\alpha)$, $\log(\mathrm{CNO})$ changes by $1.03$ and $\log(\mathrm{Fe})$ by $1.13$.

In terms of temperature, the results unsurprisingly reveal that the bulk of the emission lines arise from the Warm Ionized Medium (WIM) at temperatures between $10^4$ and $10^5$~K. The next most important contributions come from colder phases at temperatures around $\sim 10^{3.5}$~K and $\sim 10^{1.3}$~K. Finally, although much fainter, a coronal phase with temperatures above $10^5$~K is also detected (see Fig. \ref{fig:CND_weights}).

The comparison between the final model spectrum and the observations for the CND is presented in Figs.~\ref{fig:CND_comparison_spectra} and \ref{fig:CND_detailed_comparison_spectra}. The overall agreement is excellent: both the strongest and most of the fainter emission lines are accurately reproduced across the full spectral range. The most prominent residual appears near 23$\mu$m and corresponds to the \ion{Fe}{iii} line. However, this discrepancy comes from a limitation in the data processing rather than a deficiency in the model: the line flux in the observed data is distributed over two adjacent spectral bins, while our model assumes unresolved lines confined to a single bin. As a result, the flux in one of these bins (the fainter one) is not accounted for, producing a residual that nevertheless remains below 2 percent of the total flux. All other residuals are an order of magnitude smaller and show no systematic trend, attesting to the robustness of the fit.

\subsubsection{CC}

For the CC, the results are globally similar: above-solar abundances are favored for the three groups of elements, with CNO and $\alpha$ elements enriched and Fe relatively depleted. The relative abundances are
\[
\log\!\left(\frac{\mathrm{CNO}}{\alpha}\right) = 0.27 \pm 0.20, \qquad
\log\!\left(\frac{\mathrm{Fe}}{\alpha}\right) = -0.78 \pm 0.20.
\]
The best-fitting model has absolute abundances
\[
\log\alpha = 1.4,\quad \log\mathrm{CNO} = 1.4,\quad \log\mathrm{Fe} = 0.4.
\]
The PCA-derived valley is
\[
\vec{P}_{\mathrm{CC}}(s) = 
\begin{bmatrix} 1.16 \\ 0.89 \\ 0.38 \end{bmatrix}
+ s \cdot 
\begin{bmatrix} 0.93 \\ 1.00 \\ 1.05 \end{bmatrix},
\]
indicating that for every unit change in $\log(\alpha)$, $\log(\mathrm{CNO})$ changes by approximately $0.93$ and $\log(\mathrm{Fe})$ by $1.05$.

The bulk of the emission again arises from the WIM phase between $10^4$ and $10^5$ K. Regarding the contribution from the various phases, one can notice that, contrary to the CND region, there is very little contribution from molecular ($T \leq 10^4$ K) and coronal ($T \geq 10^5$ K) phases in the CC. The absence of a molecular phase is expected, as the CC is known to be depleted of molecular gas; however, the weakness of the hot phase is more surprising given the highly energetic environment. This point is discussed in more detail in Section \ref{sec:disc}. 

The model–observation comparison is shown in Figs.~\ref{fig:CC_comparison_spectra} and \ref{fig:CC_detailed_comparison_spectra}. The quality of the fit remains high, although a few features exhibit slightly stronger residuals than in the CND. The same unresolved line binning issue affects the \ion{S}{iii} line at $\sim$18$\mu$m and, to a lesser extent, the \ion{Fe}{iii} line. Additionally, two emission features show genuine mismatches between model and observation: the \ion{Na}{iii} line at $\sim$7.3~$\mu$m is significantly underpredicted by the model, while the complex blend near 14.4~$\mu$m--comprising \ion{Cl}{ii}, \ion{Na}{vi}, and \ion{Ne}{v}--is slightly overpredicted. These specific mismatches are discussed further in Section~\ref{sec:disc}. Outside of these exceptions, the agreement between the model and the data is excellent across the rest of the spectrum.

\begin{table}
    \centering
    \caption{Abundance analysis}
    \begin{tabular}{lccc}
        \hline
        \hline
        & CC & CND \\
        \hline
        log($\alpha$/$\alpha_\odot$) & $1.4$ & $2$ \\
        log(CNO/CNO$_\odot$) & $1.4$ & $1.8$ \\
        log(Fe/Fe$_\odot$) & $0.4$ & $ 1.2$ \\
        \hline
        log(CNO/$\alpha$) & $0.27 \pm 0.20$ & $ 0.05 \pm 0.26$ \\
        log(Fe/$\alpha$) & $-0.78 \pm 0.20$ & $-0.84 \pm 0.26$ \\
    \hline
    \end{tabular}
    \tablefoot{Abundance analysis from multiphase fitting. Abundances are reported as $\log_{10}(X/X_\odot)$ and ratios as $\log_{10}\!\big[(X/X_\odot)/(Y/Y_\odot)\big]$ (see notation at the start of Section~\ref{subsec:composite_and_abundance}). For each region, we report the absolute abundances of the best-fitting model, and the optimal abundance ratios as determined with the PCA analysis.}
    \label{tab:abundances}
\end{table}

\subsection{Spatial distribution of the multiphase medium}
\label{subsec:spatial}

Using the composite spectral templates derived from the integrated best-fit model, we mapped the spatial distribution of the four gas phases across the observed fields. For each pixel, the observed spectrum was decomposed into a linear combination of the molecular, warm ionized, warm coronal, and hot coronal gas components, as defined by their temperature ranges.

The resulting maps of phase contributions are presented in Figs.~\ref{fig:CND_spatial} and \ref{fig:CC_spatial} for the CND and CC regions, respectively. The coronal phase maps (right panels) were smoothed with a $2 \times 2$ Gaussian kernel to reduce noise due to the low signal level of this component. All maps are normalized to the total integrated flux of the average spectrum.

In the CND (Fig. \ref{fig:CND_spatial}), the molecular gas (left) emission is strong and exhibits a non-uniform distribution, with higher contributions in the northwestern region near the edge of the field of view and in the center, where a large cloud contributes significantly to the flux; this feature likely corresponds to the molecular “Triop” identified by \citet{Moser2017}. The warm ionized gas (middle) is very strong and almost ubiquitous, with a local minimum at the position of the aforementioned molecular cloud. More globally, the warm ionized and molecular phases appear anti-correlated. The warm coronal phase is strong and diffuse, but exhibits a maximum of emission at the interface between the molecular cloud and the peak of the warm ionized emission. The hot coronal phase is faint but displays a noticeable elongated structure oriented approximately along the north-south direction, orthogonal to the position of Sgr~A*.

In the CC (Fig. \ref{fig:CC_spatial}), the molecular contribution is much weaker and only present in a few localized patches within the field of view. In contrast, the warm ionized phase clearly dominates the flux throughout the region, although it appears noticeably more structured than in the CND, with localized enhancements and depletions forming a patchier distribution. The warm coronal phase exhibits a filamentary structure marking the edge of the warm ionized medium. Finally, the hot coronal phase is distributed in faint patches, without a clear elongated structure as observed in the CND.

\subsection{Line identifications}
\label{subsec:line_id}
As was discussed in Section \ref{subsubsec:multiphas}, our spectral fitting procedure does not assume any identification for the detected emission lines. This is a key strength of our method, as the high density of emission lines -- combined with possible line blending and Doppler shifts -- makes line identification unreliable. However, once the best-fit model is obtained, it can be used to provide a physically motivated identification of the observed emission features.

Tables~\ref{tab:line_identificationcnd} and \ref{tab:line_identificationcc} present the results of this model-based line identification for the CND and CC, respectively. For each region, we list the 20 strongest emission features and the associated lines and their predicted fluxes together with the total measured flux in the observed integrated spectrum.

Most spectral features are well identified with a single dominant transition. However, as is expected in such a crowded spectral environment, the model reveals that several observed lines arise from the blending of multiple emission lines. For instance, the feature at 5.452–5.584~$\mu$m in the CND is composed of at least three H$_2$ transitions and a He~II line, while the complex at 14.316–14.448~$\mu$m in the CC includes contributions from [Cl~II], [Na~VI], and [Ne~V]. These blends illustrate the value of model-based identification.

In most cases, the agreement between the model-predicted and observed fluxes is good, such as for the strong [Ne~II] line at 12.81~$\mu$m in both regions. However, a few lines--such as [Ni~II] at 6.63~$\mu$m in the CC--exhibit significant deviations, possibly indicating limitations of our modeling.

We make the full prediction of our model for the two regions publicly available. They are downloadable as ASCII table from the CDS.

\begin{table*}
\caption{Brightest emission lines in the CND}
\centering
\begin{tabular}{cccc}
\hline
\hline
Lines & Wavelengths ($\mu$ m) & Model flux (W/m$^2$/sr) & Measured flux (W/m$^2$/sr) \\
\hline
 &5.298 - 5.386 & 7.33 $\times$ 10$^{-7}$ & 5.10 $\times$ 10$^{-7}$\ $\pm$ \ 5.00 $\times$ 10$^{-9}$\\
\ion{Fe}{ii} & 5.339 & 7.33 $\times$ 10$^{-7}$ & - \\
\hline
 &5.474 - 5.562 & 1.31 $\times$ 10$^{-6}$ & 6.85 $\times$ 10$^{-7}$\ $\pm$ \ 9.73 $\times$ 10$^{-9}$\\
H$_2$ & 5.510 & 1.22 $\times$ 10$^{-6}$ & - \\
H$_2$ & 5.554 & 8.60 $\times$ 10$^{-8}$ & - \\
\hline
 &6.091 - 6.180 & 4.74 $\times$ 10$^{-7}$ & 2.33 $\times$ 10$^{-7}$\ $\pm$ \ 2.00 $\times$ 10$^{-9}$\\
H$_2$ & 6.107 & 4.67 $\times$ 10$^{-7}$ & - \\
H$_2$ & 6.101 & 6.68 $\times$ 10$^{-9}$ & - \\
\hline
 &6.621 - 6.709 & 4.28 $\times$ 10$^{-7}$ & 1.36 $\times$ 10$^{-7}$\ $\pm$ \ 1.42 $\times$ 10$^{-9}$\\
\ion{Ni}{ii} & 6.634 & 4.28 $\times$ 10$^{-7}$ & - \\
\hline
 &6.885 - 6.973 & 1.51 $\times$ 10$^{-6}$ & 1.11 $\times$ 10$^{-6}$\ $\pm$ \ 9.20 $\times$ 10$^{-9}$\\
\ion{Ni}{ii} & 6.918 & 1.51 $\times$ 10$^{-6}$ & - \\
\hline
 &6.973 - 7.062 & 1.00 $\times$ 10$^{-5}$ & 6.53 $\times$ 10$^{-6}$\ $\pm$ \ 4.97 $\times$ 10$^{-8}$\\
\ion{Ar}{ii} & 6.983 & 1.00 $\times$ 10$^{-5}$ & - \\
\hline
 &7.988 - 8.076 & 5.86 $\times$ 10$^{-7}$ & 4.78 $\times$ 10$^{-7}$\ $\pm$ \ 2.89 $\times$ 10$^{-9}$\\
H$_2$ & 8.023 & 5.86 $\times$ 10$^{-7}$ & - \\
\hline
 &8.958 - 9.046 & 2.50 $\times$ 10$^{-7}$ & 6.94 $\times$ 10$^{-8}$\ $\pm$ \ 4.84 $\times$ 10$^{-10}$\\
\ion{Ar}{iii} & 8.989 & 2.50 $\times$ 10$^{-7}$ & - \\
\hline
 &12.265 - 12.354 & 2.71 $\times$ 10$^{-7}$ & 2.23 $\times$ 10$^{-7}$\ $\pm$ \ 1.84 $\times$ 10$^{-9}$\\
\ion{Fe}{vi} & 12.307 & 2.10 $\times$ 10$^{-7}$ & - \\
H$_2$ & 12.275 & 6.06 $\times$ 10$^{-8}$ & - \\
\hline
 &12.795 - 12.883 & 1.58 $\times$ 10$^{-5}$ & 1.60 $\times$ 10$^{-5}$\ $\pm$ \ 1.70 $\times$ 10$^{-7}$\\
\ion{Ne}{ii} & 12.810 & 1.58 $\times$ 10$^{-5}$ & - \\
\hline
 &14.338 - 14.426 & 1.97 $\times$ 10$^{-7}$ & 1.36 $\times$ 10$^{-7}$\ $\pm$ \ 1.41 $\times$ 10$^{-9}$\\
\ion{Cl}{ii} & 14.364 & 1.97 $\times$ 10$^{-7}$ & - \\
\hline
 &15.529 - 15.617 & 6.77 $\times$ 10$^{-7}$ & 6.46 $\times$ 10$^{-7}$\ $\pm$ \ 5.81 $\times$ 10$^{-9}$\\
\ion{Ne}{iii} & 15.551 & 6.77 $\times$ 10$^{-7}$ & - \\
\hline
 &17.028 - 17.116 & 1.66 $\times$ 10$^{-7}$ & 1.36 $\times$ 10$^{-7}$\ $\pm$ \ 1.12 $\times$ 10$^{-9}$\\
H$_2$ & 17.030 & 1.66 $\times$ 10$^{-7}$ & - \\
\hline
 &17.910 - 17.998 & 1.16 $\times$ 10$^{-7}$ & 1.15 $\times$ 10$^{-7}$\ $\pm$ \ 1.08 $\times$ 10$^{-9}$\\
\ion{Fe}{ii} & 17.931 & 1.16 $\times$ 10$^{-7}$ & - \\
\hline
 &18.704 - 18.792 & 2.29 $\times$ 10$^{-6}$ & 2.79 $\times$ 10$^{-6}$\ $\pm$ \ 2.43 $\times$ 10$^{-8}$\\
\ion{S}{iii} & 18.708 & 2.29 $\times$ 10$^{-6}$ & - \\
\hline
 &21.791 - 21.879 & 3.59 $\times$ 10$^{-8}$ & 2.56 $\times$ 10$^{-8}$\ $\pm$ \ 1.69 $\times$ 10$^{-10}$\\
\ion{Ar}{iii} & 21.825 & 3.59 $\times$ 10$^{-8}$ & - \\
\hline
 &22.893 - 22.981 & 6.21 $\times$ 10$^{-7}$ & 8.98 $\times$ 10$^{-7}$\ $\pm$ \ 8.23 $\times$ 10$^{-9}$\\
\ion{Fe}{iii} & 22.919 & 6.21 $\times$ 10$^{-7}$ & - \\
\hline
 &24.481 - 24.569 & 4.37 $\times$ 10$^{-8}$ & 4.03 $\times$ 10$^{-8}$\ $\pm$ \ 4.40 $\times$ 10$^{-10}$\\
\ion{Fe}{ii} & 24.513 & 4.37 $\times$ 10$^{-8}$ & - \\
\hline
 &25.231 - 25.319 & 5.01 $\times$ 10$^{-8}$ & 4.39 $\times$ 10$^{-8}$\ $\pm$ \ 3.43 $\times$ 10$^{-10}$\\
\ion{S}{i} & 25.242 & 5.01 $\times$ 10$^{-8}$ & - \\
\hline
 &25.980 - 26.068 & 7.11 $\times$ 10$^{-7}$ & 7.40 $\times$ 10$^{-7}$\ $\pm$ \ 9.39 $\times$ 10$^{-9}$\\
\ion{Fe}{ii} & 25.981 & 7.11 $\times$ 10$^{-7}$ & - \\
\hline
\end{tabular}
\tablefoot{List of the 20 brightest emission features detected in the CND spectrum, sorted by wavelength. For each feature, the wavelength interval, total flux from the model, and measured flux are listed in the top row. The subsequent rows indicate the individual emission lines predicted by the model within that interval, including their rest wavelengths and modeled fluxes. When several lines fall within the same spectral bin, they are all listed, illustrating the possible line blending. The modeled flux values are those from the best-fit multiphase model.}
\label{tab:line_identificationcnd}
\end{table*}

\begin{table*}
\caption{Same as Table \ref{tab:line_identificationcnd}, but for the CC}
\centering
\begin{tabular}{cccc}
\hline
\hline
Lines & Wavelengths ($\mu$ m) & Model flux (W/m$^2$/sr) & Measured flux (W/m$^2$/sr) \\
\hline
 &5.298 - 5.386 & 1.67 $\times$ 10$^{-6}$ & 9.22 $\times$ 10$^{-7}$\ $\pm$ \ 6.01 $\times$ 10$^{-9}$\\
\ion{Ni}{ix} & 5.318 & 1.61 $\times$ 10$^{-6}$ & - \\
\ion{Fe}{ii} & 5.339 & 5.12 $\times$ 10$^{-8}$ & - \\
\hline
 &6.621 - 6.709 & 3.55 $\times$ 10$^{-6}$ & 7.24 $\times$ 10$^{-7}$\ $\pm$ \ 7.81 $\times$ 10$^{-9}$\\
\ion{Ni}{ii} & 6.634 & 3.55 $\times$ 10$^{-6}$ & - \\
\hline
 &6.973 - 7.062 & 5.60 $\times$ 10$^{-5}$ & 3.82 $\times$ 10$^{-5}$\ $\pm$ \ 2.72 $\times$ 10$^{-7}$\\
\ion{Ar}{ii} & 6.983 & 5.60 $\times$ 10$^{-5}$ & - \\
\hline
 &7.414 - 7.503 & 2.92 $\times$ 10$^{-6}$ & 1.83 $\times$ 10$^{-6}$\ $\pm$ \ 1.29 $\times$ 10$^{-8}$\\
\ion{H}{i} & 7.458 & 2.58 $\times$ 10$^{-6}$ & - \\
\ion{H}{i} & 7.500 & 3.33 $\times$ 10$^{-7}$ & - \\
\hline
 &8.958 - 9.046 & 1.31 $\times$ 10$^{-6}$ & 5.94 $\times$ 10$^{-7}$\ $\pm$ \ 4.68 $\times$ 10$^{-9}$\\
\ion{Mg}{vii} & 9.031 & 1.02 $\times$ 10$^{-6}$ & - \\
\ion{Ar}{iii} & 8.989 & 2.94 $\times$ 10$^{-7}$ & - \\
\hline
 &10.501 - 10.590 & 4.78 $\times$ 10$^{-7}$ & -1.99 $\times$ 10$^{-8}$\ $\pm$ \ 2.49 $\times$ 10$^{-10}$\\
\ion{S}{iv} & 10.508 & 4.78 $\times$ 10$^{-7}$ & - \\
\hline
 &12.354 - 12.442 & 7.11 $\times$ 10$^{-7}$ & 4.46 $\times$ 10$^{-7}$\ $\pm$ \ 3.78 $\times$ 10$^{-9}$\\
\ion{H}{i} & 12.368 & 7.11 $\times$ 10$^{-7}$ & - \\
\hline
 &12.706 - 12.795 & 3.68 $\times$ 10$^{-7}$ & 1.78 $\times$ 10$^{-7}$\ $\pm$ \ 1.66 $\times$ 10$^{-9}$\\
\ion{Ni}{ii} & 12.725 & 3.68 $\times$ 10$^{-7}$ & - \\
\hline
 &12.795 - 12.883 & 4.16 $\times$ 10$^{-5}$ & 3.96 $\times$ 10$^{-5}$\ $\pm$ \ 3.71 $\times$ 10$^{-7}$\\
\ion{Ne}{ii} & 12.810 & 4.16 $\times$ 10$^{-5}$ & - \\
\hline
 &14.338 - 14.426 & 1.09 $\times$ 10$^{-6}$ & 2.48 $\times$ 10$^{-7}$\ $\pm$ \ 1.71 $\times$ 10$^{-9}$\\
\ion{Cl}{ii} & 14.364 & 8.76 $\times$ 10$^{-7}$ & - \\
\ion{Na}{vi} & 14.393 & 2.12 $\times$ 10$^{-7}$ & - \\
\hline
 &15.529 - 15.617 & 4.48 $\times$ 10$^{-6}$ & 5.28 $\times$ 10$^{-6}$\ $\pm$ \ 5.30 $\times$ 10$^{-8}$\\
\ion{Ne}{iii} & 15.551 & 4.48 $\times$ 10$^{-6}$ & - \\
\hline
 &17.910 - 17.998 & 3.15 $\times$ 10$^{-7}$ & 2.68 $\times$ 10$^{-7}$\ $\pm$ \ 2.04 $\times$ 10$^{-9}$\\
\ion{Fe}{ii} & 17.931 & 3.15 $\times$ 10$^{-7}$ & - \\
\hline
 &18.704 - 18.792 & 7.62 $\times$ 10$^{-6}$ & 9.44 $\times$ 10$^{-6}$\ $\pm$ \ 7.60 $\times$ 10$^{-8}$\\
\ion{S}{iii} & 18.708 & 7.62 $\times$ 10$^{-6}$ & - \\
\hline
 &21.791 - 21.879 & 2.00 $\times$ 10$^{-7}$ & 1.87 $\times$ 10$^{-7}$\ $\pm$ \ 1.66 $\times$ 10$^{-9}$\\
\ion{Ar}{iii} & 21.825 & 2.00 $\times$ 10$^{-7}$ & - \\
\hline
 &22.893 - 22.981 & 5.61 $\times$ 10$^{-6}$ & 6.03 $\times$ 10$^{-6}$\ $\pm$ \ 6.37 $\times$ 10$^{-8}$\\
\ion{Fe}{iii} & 22.919 & 5.61 $\times$ 10$^{-6}$ & - \\
\hline
 &24.304 - 24.393 & 1.08 $\times$ 10$^{-7}$ & 2.48 $\times$ 10$^{-7}$\ $\pm$ \ 2.90 $\times$ 10$^{-9}$\\
\ion{Ne}{v} & 24.311 & 1.08 $\times$ 10$^{-7}$ & - \\
\hline
 &24.481 - 24.569 & 1.56 $\times$ 10$^{-7}$ & 1.15 $\times$ 10$^{-7}$\ $\pm$ \ 9.02 $\times$ 10$^{-10}$\\
\ion{Fe}{ii} & 24.513 & 1.56 $\times$ 10$^{-7}$ & - \\
\hline
 &25.848 - 25.936 & 1.86 $\times$ 10$^{-7}$ & 1.68 $\times$ 10$^{-7}$\ $\pm$ \ 1.78 $\times$ 10$^{-9}$\\
\ion{O}{iv} & 25.886 & 1.77 $\times$ 10$^{-7}$ & - \\
\ion{Fe}{v} & 25.913 & 9.15 $\times$ 10$^{-9}$ & - \\
\hline
 &25.892 - 25.980 & 1.53 $\times$ 10$^{-7}$ & 2.66 $\times$ 10$^{-7}$\ $\pm$ \ 2.98 $\times$ 10$^{-9}$\\
\ion{Fe}{v} & 25.913 & 1.32 $\times$ 10$^{-7}$ & - \\
H$_2$O & 25.933 & 1.58 $\times$ 10$^{-8}$ & - \\
H$_2$O & 25.978 & 5.25 $\times$ 10$^{-9}$ & - \\
\hline
 &25.980 - 26.068 & 7.59 $\times$ 10$^{-7}$ & 8.99 $\times$ 10$^{-7}$\ $\pm$ \ 9.68 $\times$ 10$^{-9}$\\
\ion{Fe}{ii} & 25.981 & 7.59 $\times$ 10$^{-7}$ & - \\
\hline
\end{tabular}
\label{tab:line_identificationcc}
\end{table*}

\section{Discussion}
\label{sec:disc}
\subsection{A multiphase ISM}
\subsubsection{Presence of ionized and coronal phase in the CND}
The presence of strong ionized and coronal emission in the CND, typically considered a dense molecular structure, points to an active and multiphase ISM. While the young massive stellar population lies in the CC, the reach of its stellar winds -- especially from Wolf-Rayet stars -- can extend up to dozens of parsec \citep{Cazzolato2000, vanMarle2004, Freyer2006} well into and beyond the CND. These winds generate shocks that heat and ionize gas, producing the observed high-excitation lines. Additionally, outflows from Sgr~A*, whether ongoing or relic, may drive large-scale shocks that traverse the CND. Another contributor could be magnetic turbulence, which is expected to be enhanced in the dense, rotating, and shearing environment of the CND due to its strong toroidal magnetic fields and differential motions \citep{Morris1996, Dinh2021, Guerra2023, Sato2024}. Dissipation of such turbulence can produce localized shock heating, producing ionized phases even in dense regions \citep{Yusef2001, Melia2001}. However, the survival of such a high temperature phase also raises questions about the shielding and survival of molecular gas in proximity to these energetic processes, as was already raised in the study of hot molecular hydrogen by \citet{Ciurlo2016}.

\subsubsection{High densities}

The maps of temperature and density contributions (Figs.~\ref{fig:CND_weights} and \ref{fig:CC_weights}) indicate that most of the emission originates from high-density gas ($n\geq10^3$~cm$^{-3}$), even for the warm and hot ionized phases, which are expected to be less dense than molecular. While this may initially appear surprising, it reflects the well-known behavior of emission line processes: line emissivities for most atomic and ionic transitions scale approximately with the square of the gas density. As a result, small volumes of dense gas can dominate the observed emission, even when embedded within a much more massive but diffuse medium.

Thus, these maps should be interpreted with care: they reflect the flux contribution, not the mass or volume fraction of the gas. The apparent dominance of high-density components is consistent with a structured, inhomogeneous medium composed of dense clumps or filaments within a more diffuse interclump environment. In particular, it is likely that low-density material dominates the mass budget in both the CND and the CC, but contributes less to the emission budget traced by our data.

In contrast, for the cold molecular phase, many of the relevant emission lines (e.g., rotational lines of H$_ 2$) do not exhibit such a steep density dependence. Therefore, the high densities inferred for the molecular component likely trace not only the peak flux contributors but also the bulk of the cold gas mass. This is consistent with independent knowledge that the molecular gas in the CND is indeed dense.

\subsection{Abundances}

\subsubsection{Star-formation induced CNO and $\alpha$ enhancement}

The observed enhancement in CNO and $\alpha$ elements in the two regions supports the presence of a relatively recent star formation episode in the GC. These elements are predominantly synthesized in massive stars and released into the ISM through core-collapse supernovae, enriching their surroundings on timescales of a few to a few tens of Myr. This is consistent with the last known major star formation event in the central parsec, dated to less than 10 Myr \citep{Krabbe1995, Feldmeier2015, Mastrobuono2019}. It is noteworthy that this conclusion arises solely from the analysis of ionized gas emission lines, independent of any direct stellar population study.

\subsubsection{Very high abundances}

Our models yield a wide range of acceptable abundance values, generally constraining the relative ratios between CNO, $\alpha$ elements, and Fe more robustly than their absolute abundances. The best-fit values for the absolute abundances are often very high, with $\log(\mathrm{CNO/H})$ or $\log(\alpha/\mathrm{H})$ typically between 1 and 2 (i.e., 10–100× solar). Such extreme values are surprising when compared with previous measurements. For instance, studies of hot gas near Sgr~A* assuming WR wind enrichment suggest near-solar or subsolar abundances depending on light-element assumptions \citep{Hua2023, Hua2025}, and X-ray studies estimate iron abundances of at most $\sim$2× solar \citep{Anastasopoulou2023}. In the stellar component, most red giant stars show metallicities close to solar, with a spread ranging from [M/H] $\sim -1.25$ to $+0.5$ dex \citep{FeldmeierKrause2020, Do2015}. Given this context, the most plausible interpretation is that our models overestimate the absolute abundances, and the true values likely lie toward the lower end of our fitted range. Nonetheless, our diagnostics confidently rule out subsolar values (see Figs. \ref{fig:CND_abundances} and \ref{fig:CC_abundances}), pointing instead to solar or mildly supersolar metallicity as the most realistic estimate.

\subsubsection{Relative Fe depletion}

The relative depletion of iron compared to $\alpha$ elements may point to two complementary effects: chemical youth and dust grain sequestration. Iron is largely produced by Type Ia supernovae, which arise from long-lived binary systems, whereas $\alpha$ elements and CNO elements originate primarily from core-collapse supernovae and AGB stars, both associated with shorter stellar lifetimes \citep{Matteucci1986, Woosley1995, Nomoto2013, Karakas2014}. A low Fe/$\alpha$ and Fe/CNO ratio is thus a signature of recent, short-timescale star formation activity, with insufficient time for significant iron enrichment from Type Ia events \citep{Centurion2000}. At the same time, Fe is highly refractory and easily incorporated from the ISM into dust grains \citep{DelgadoInglada2009, Gioannini2016}, which are known to be abundant in the CND \citep{Latvakoski1999, Etxaluze2011}. This dual interpretation--chemical youth and dust sequestration--supports the picture of a young and dusty environment at the GC and is not consistent with a gradual accumulation of metals from successive star formation cycles over galactic timescales.

\subsection{Spatial variations}

In the CND, molecular and warm ionized gas exhibit a clear spatial complementarity, with ionized emission often wrapping around or bordering molecular structures. This suggests a layered ISM, likely shaped by UV radiation and stellar winds forming photo-dissociation regions, or by shocks compressing molecular gas while ionizing its surroundings.

The spatial morphology of high-temperature coronal line emission in the CND reveals a prominent, North-South oriented filamentary structure, which contrasts with the surrounding gas geometry. These features trace plasma at temperatures exceeding $10^6$ K. Its orientation -- roughly perpendicular to the line connecting its position to Sgr~A* -- points to the action of a large-scale shock propagating through the region. Such a shock could result from past energetic outflows driven by the central SMBH or from the collective impact of stellar winds from the nearby young stellar cluster. 

Emission from intermediate-temperature gas ($10^{4.7}$ - $10^5$ K) appears mainly at the boundaries between molecular clouds and the surrounding warm ionized medium. This phase is especially prominent at cloud interfaces in both the CND and CC. In the CND, the bulk of emission from this phase appears at the North-West edge of the largest molecular structure, spatially located between the molecular and warm ionized phase, suggesting a transition zone due to shocks or photoionization. In the CC, where molecular tracers are weaker, the ionized phase still appears to follow the path of the minispiral arms, supporting a similar interface-based interpretation.

\section{Conclusions}

We present a detailed spectroscopic analysis of JWST MIRI/MRS observations. It focuses on two regions in the GC: the CND and the CC.

The first part of our analysis focuses on the integrated spectra of these regions. We fit these spectra using a multiphase CLOUDY model spanning a wide range of densities, temperatures, and elemental abundances.

We find that, in both regions, the emission-line spectrum is dominated in flux by the warm ionized phase of the gas, though contributions from molecular and hot coronal phases are also present. As expected, the molecular phase is more prominent in the CND. Surprisingly, the hot coronal phase, while faint, is clearly detected in both the CND and CC.

Our abundance analysis focuses on three groups of elements: the CNO group, the $\alpha$ elements, and the iron-peak elements. The relative abundances are robustly constrained: we find a clear depletion of iron-peaked elements relative to $\alpha$ with $\log(\mathrm{Fe}/\alpha) = -0.78 \pm 0.20$ (CC) and $-0.84 \pm 0.26$ (CND), while CNO is only mildly enhanced relative to $\alpha$, with $\log(\mathrm{CNO}/\alpha) = 0.27 \pm 0.20$ (CC) and $0.05 \pm 0.26$ (CND). Absolute abundances, expressed as solar-normalized logarithms, are supersolar but more degenerate; the best-fitting models yield $(\log\alpha,\log\mathrm{CNO},\log\mathrm{Fe})=(1.4,1.4,0.4)$ in the CC and $(2.0,1.8,1.2)$ in the CND. This pattern indicates a recent or ongoing episode of star formation, and a relatively limited contribution from long-timescale enrichment processes such as Type~Ia supernovae.

In the final part of our study, we perform a spatially resolved analysis of the four principal phases of the ISM: molecular, warm ionized, hot ionized, and coronal. We observe a clear spatial complementarity between these phases, suggesting a stratified ISM structure.

\section{Data availability}
The gas-phase parameter maps for the CND and CC (Figs. \ref{fig:CND_spatial} and \ref{fig:CC_spatial}), together with the extracted spectra at native spectral resolution, the rebinned spectra used in this work, and the corresponding best-fit model spectra, are available in electronic form at the CDS via anonymous ftp to cdsarc.u-strasbg.fr (130.79.128.5) or via http://cdsweb.u-strasbg.fr/cgi-bin/qcat?J/A+A/.

\begin{acknowledgements}
The authors thank Thibaut Paumard and Yann Clénet for the support in creating and managing this project and for the insightful discussions pushing it forward.  This work was supported by French government through the National Research Agency (ANR) with funding grants ANR-21-CE31-0011 and ANR-22-EXOR-0001, and by the Thematic Action “Physique et Chimie du Milieu Interstellaire” (PCMI) of INSU Programme National “Astro”, with contributions from CNRS Physique \& CNRS Chimie, CEA, and CNES.
This work was granted access to the HPC resources of MesoPSL financed by the Region Ile de France and the project Equip@Meso (reference ANR-10-EQPX-29-01) of the programme Investissements d’Avenir supervised by the Agence Nationale pour la Recherche.
AC, MRM and JQ acknowledge support from STScI grant JWST-GO-03166.001-A. 
\end{acknowledgements}

\bibliographystyle{aa}
\bibliography{biblio}

\appendix

\section{Single temperature and density fit}
\label{app:appendix_residuals}

Before adopting the full multiphase modeling approach, we conducted a simpler fit assuming that each region could be represented by a single gas phase. For this purpose, we computed the squared residuals between each synthetic model and the observed line fluxes, using the one-to-one residual metric described in Section~\ref{sec:modeling}.

\begin{figure}
    \centering
 
    \includegraphics[width=\linewidth]{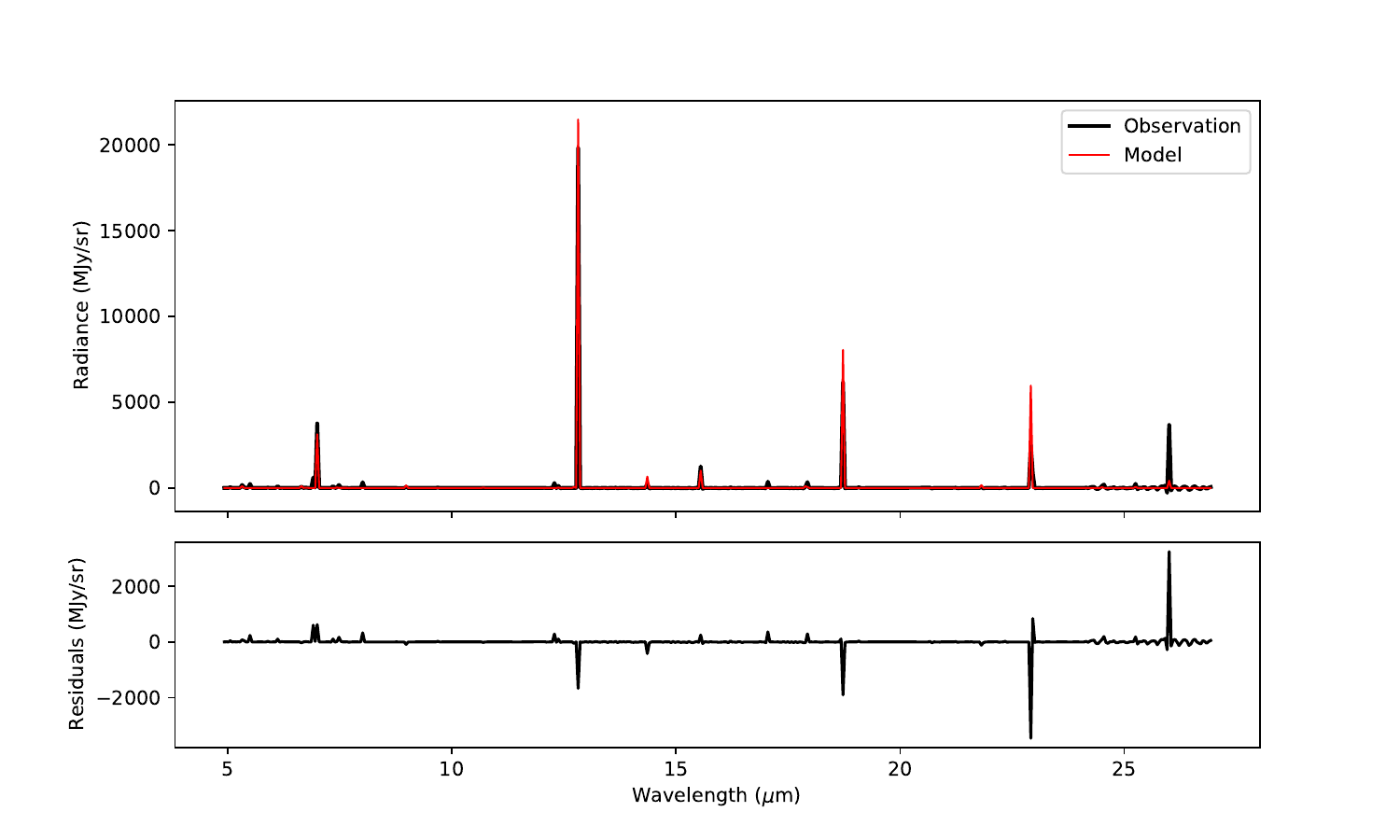}
    \caption{CND: Same as Fig.~\ref{fig:CND_comparison_spectra}, but for the best model obtained with the one-to-one matching method described in \ref{subsubsec:monophas}.}
    \label{fig:CND_residuals_comparison}
\end{figure}

\begin{figure}
    \centering
 
    \includegraphics[width=\linewidth]{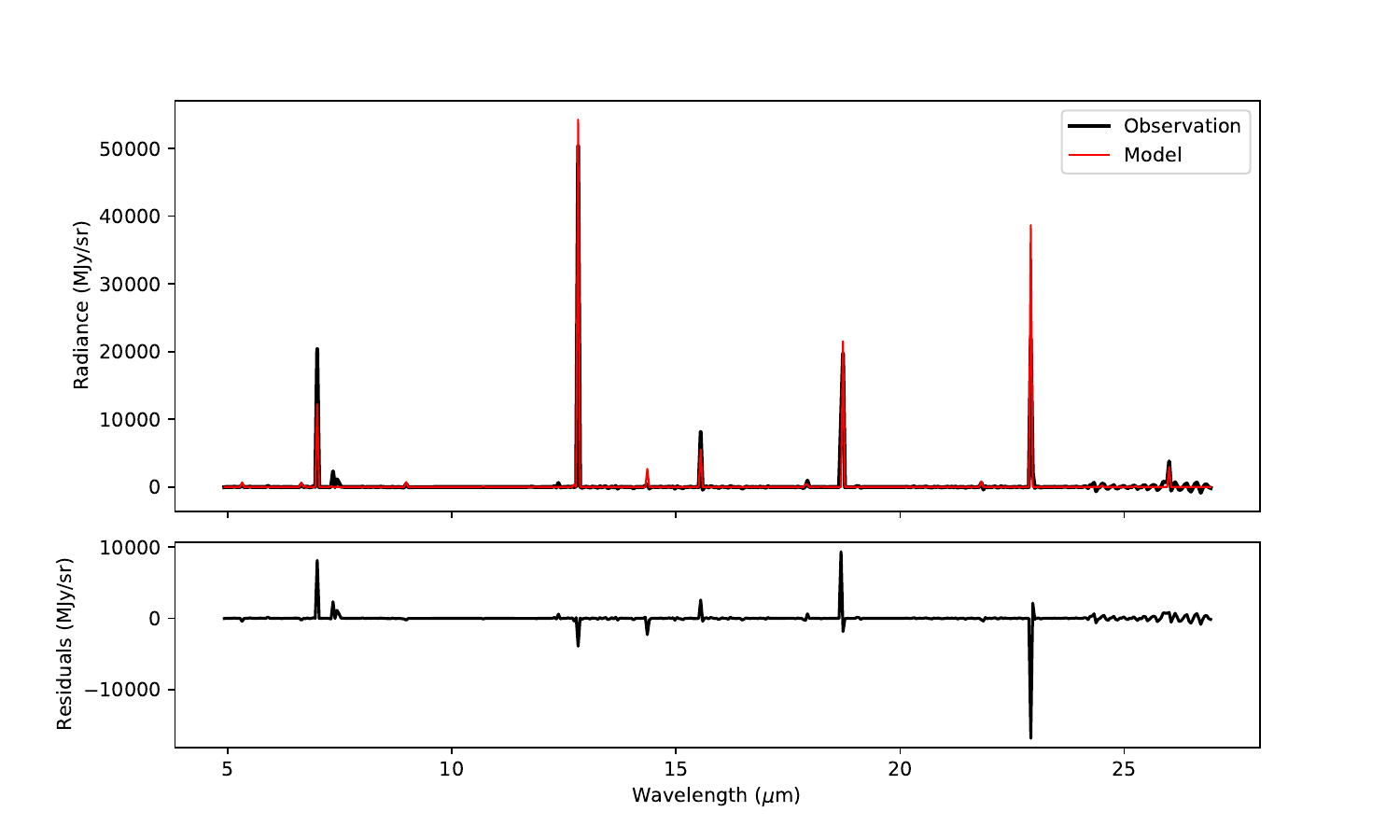}
    \caption{CC: Same as Fig.~\ref{fig:CC_comparison_spectra}, but for the best model obtained with the one-to-one matching method described in \ref{subsubsec:monophas}.}
    \label{fig:CC_residuals_comparison}
\end{figure}

Figures~\ref{fig:CND_residuals_comparison} and \ref{fig:CC_residuals_comparison} show the residual maps for the CND and CC, respectively, as functions of gas temperature and density. Although the absolute residual values are significantly higher than in the multiphase case (typically an order of magnitude worse), these maps still reveal the physical conditions that best match the spectra under this simplified assumption.

For both regions, the optimal solution corresponds to warm ionized gas with a temperature of $\sim 10^{4.3}$ K, and loose constraints on density. A secondary minimum appears at cooler temperatures ($10^3$–$10^4$ K), more pronounced in the CND, suggesting a stronger molecular contribution in that region. Despite their simplicity, these single-phase fits reveal that the bulk of the emission in both regions originates from warm ionized gas.

\begin{figure}
    \centering
    \includegraphics[width=\linewidth]{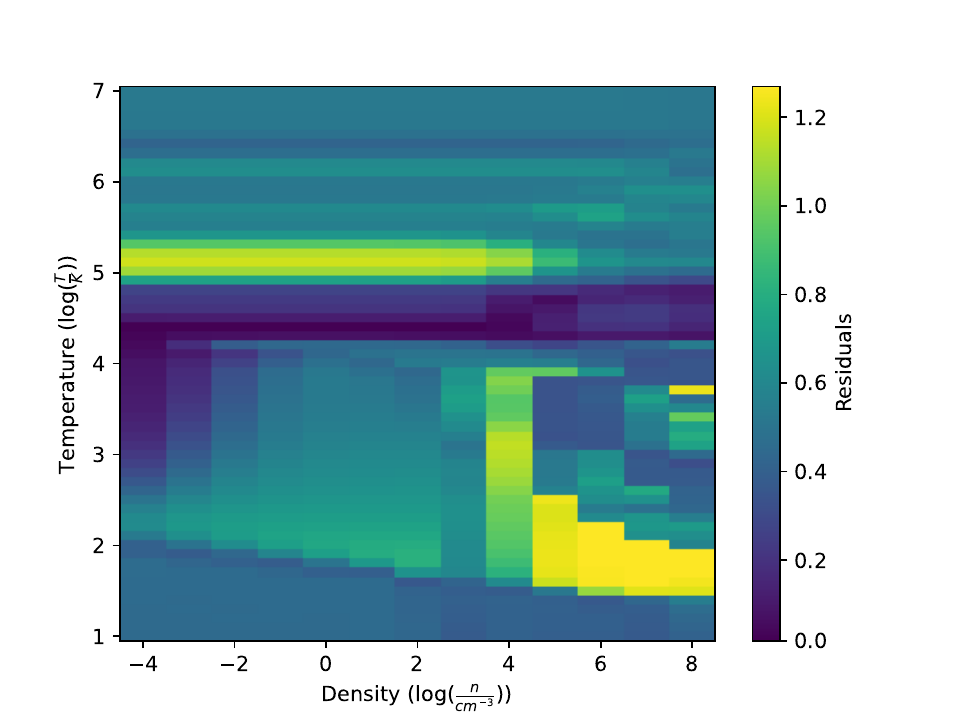}
    \caption{Residuals between the observed spectrum of the CND and individual CLOUDY simulations, plotted as a function of gas temperature (y axis) and density (x axis), for the best-fit abundance triplet identified in Table~\ref{tab:abundances}. The residuals are computed as described in Sub-section~\ref{subsubsec:monophas}, with darker colors indicating a better match. A well-defined minimum is found around \( T \sim 10^{4.4}~\mathrm{K} \), with weaker constraints on the density. Secondary minima are found at lower temperatures.}
    \label{fig:CND_residuals}
\end{figure}

\begin{figure}
    \centering
    \includegraphics[width=\linewidth]{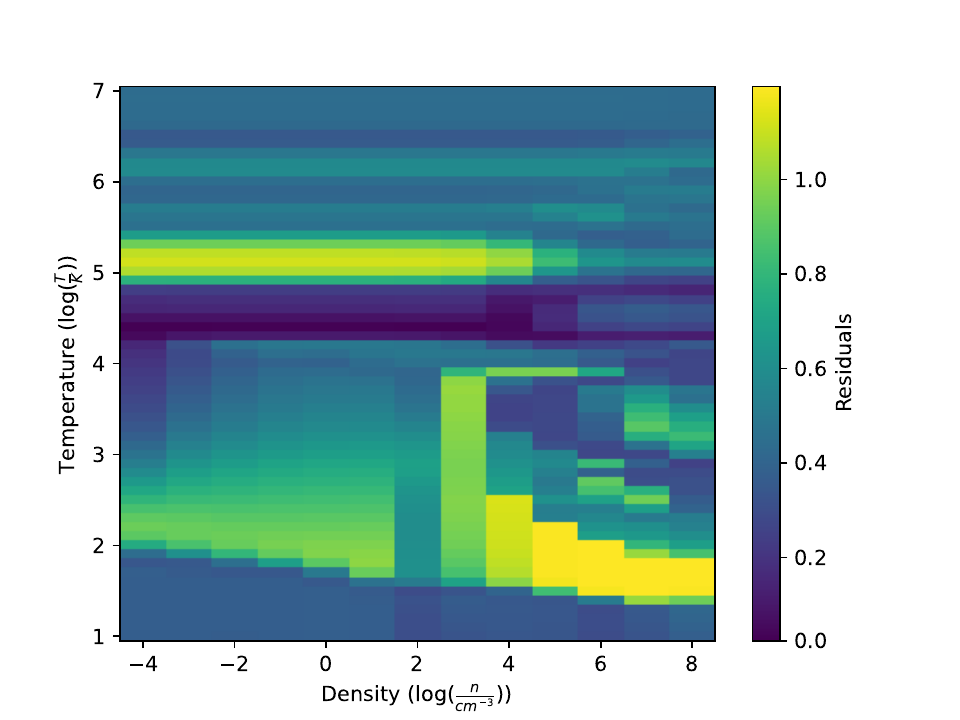}
    \caption{Same as Fig.~\ref{fig:CND_residuals}, but for the CC.}
    \label{fig:CC_residuals}
\end{figure}

\section{Multiphase medium}
\label{app:appendix_multi}

\begin{figure*}[t]
    \centering
    \includegraphics[width=0.85\linewidth]{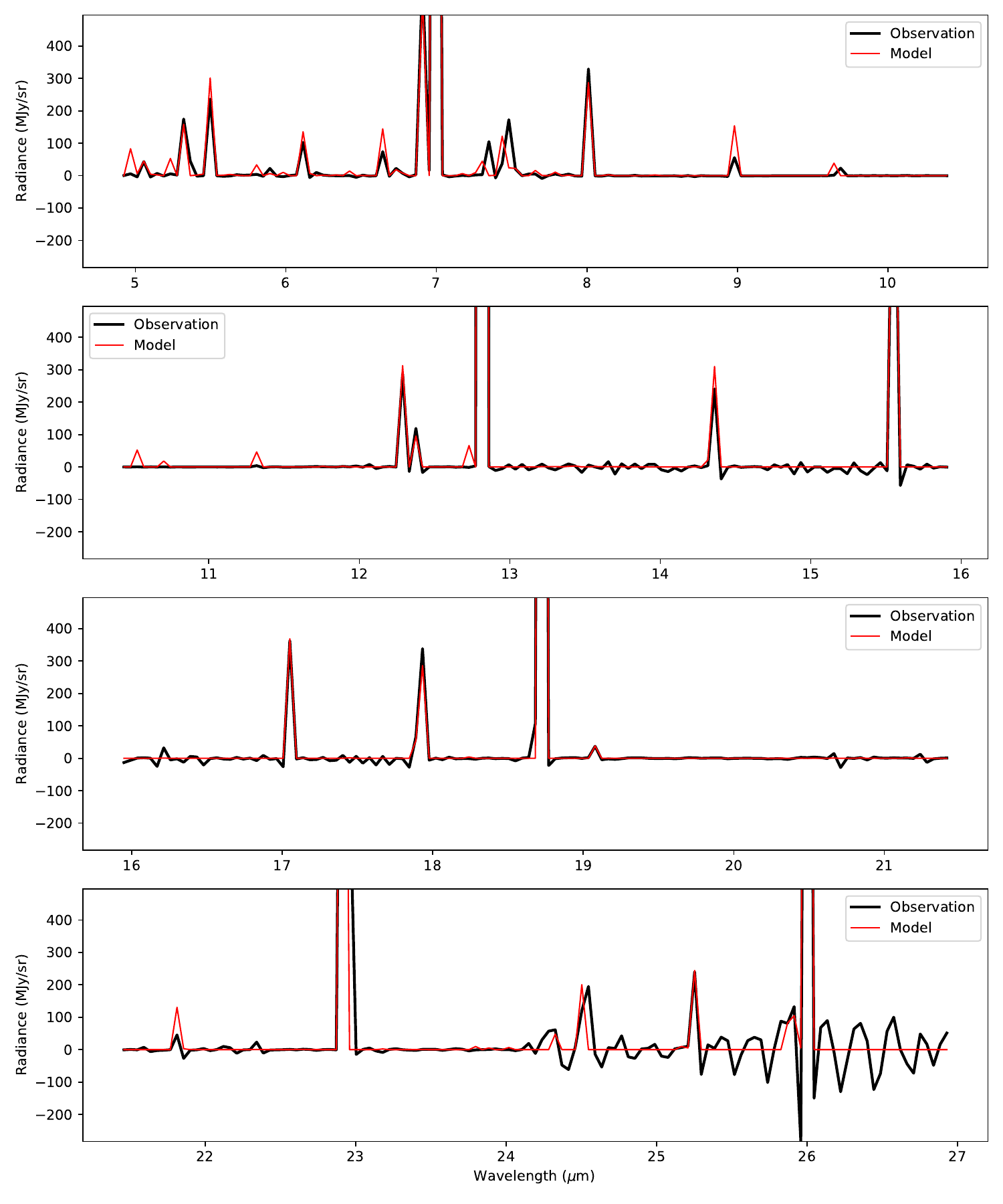}
    \caption{Zoomed-in view of Fig.~\ref{fig:CND_comparison_spectra}, showing the comparison between the observed (thick black) and modeled (thin red) spectra for the CND. The full spectral range is divided into four panels, increasing in wavelength from top to bottom. The y axis has been zoomed by a factor of 40 to emphasize weaker emission lines and differences between the model and the data.}
    \label{fig:CND_detailed_comparison_spectra}
\end{figure*}

\begin{figure*}[t]
    \centering
    \includegraphics[width=0.85\linewidth]{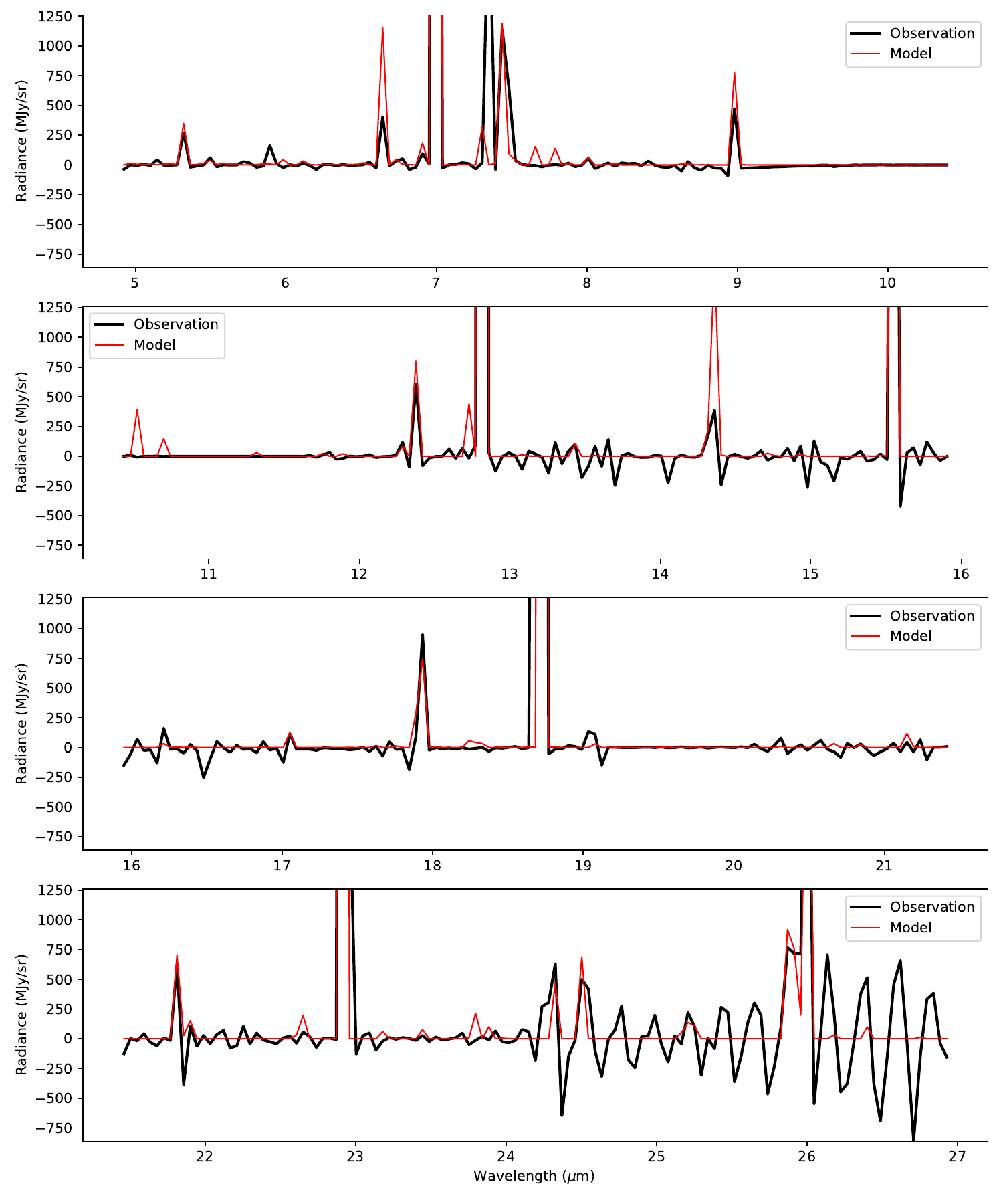}
    \caption{Zoomed-in view of Fig.~\ref{fig:CC_comparison_spectra}, showing the comparison between the observed (thick black) and modeled (thin red) spectra for the CND. The full spectral range is divided into four panels, increasing in wavelength from top to bottom. The y axis has been zoomed by a factor of 40 to emphasize weaker emission lines and differences between the model and the data.}
    \label{fig:CC_detailed_comparison_spectra}
\end{figure*}

\section{Principal Component Analysis}
\label{app:PCA}

\begin{figure*}
    \centering
    \includegraphics[width=0.95\linewidth]{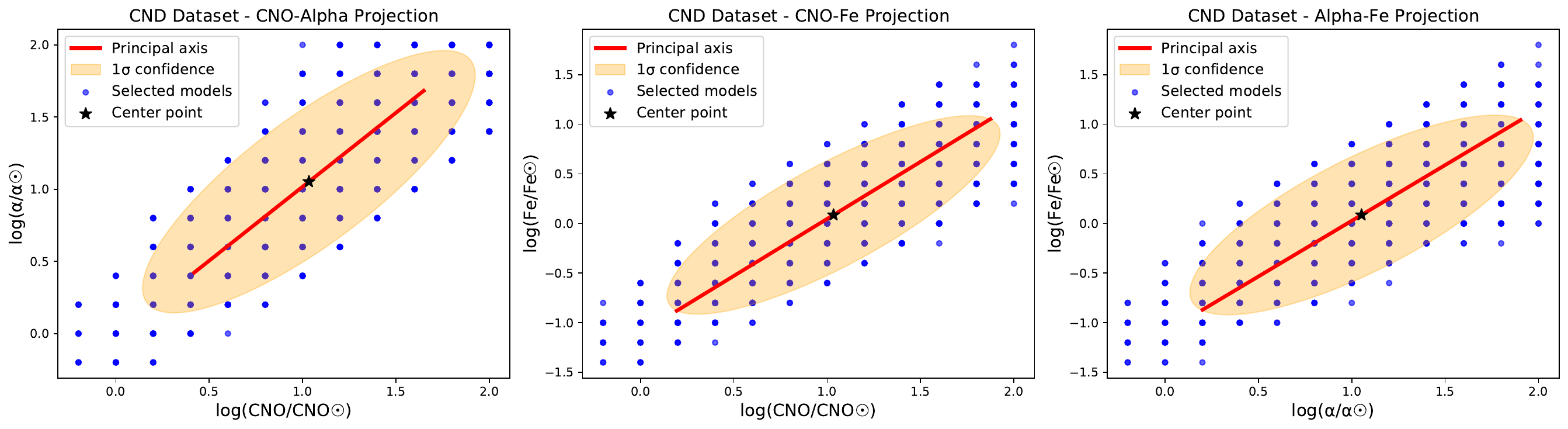}
    \caption{2D visualizations of the PCA used to determine the abundances and associated uncertainties for the CND.}
    \label{fig:CND_PCA}
\end{figure*}

\begin{figure*}
    \centering
    \includegraphics[width=0.95\linewidth]{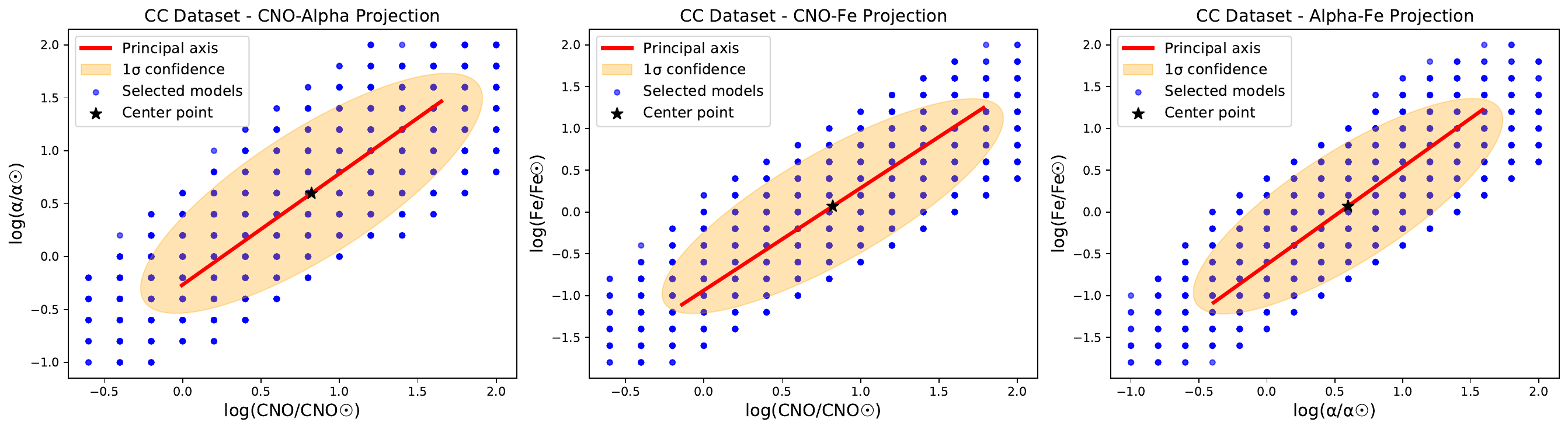}
    \caption{2D visualizations of the PCA used to determine the abundances and associated uncertainties for the CC.}
    \label{fig:CC_PCA}
\end{figure*}

\end{document}